\documentclass{PoS}

\usepackage{graphicx}
\usepackage{multirow}
\usepackage{arydshln}
\usepackage{slashed}

\def\OMIT#1{{}}
\def\lsim{\mathrel{\!\mathpalette\vereq<}\!}
\def\gsim{\mathrel{\!\mathpalette\vereq>}\!}
\def\vereq#1#2{\lower3.5pt\vbox{\baselineskip1.5pt \lineskip1.5pt
\ialign{$#1\hfill##\hfil$\crcr#2\crcr\sim\crcr}}}

\newcommand{\beq}{\begin{equation}}
\newcommand{\eeq}{\end{equation}}
\newcommand{\beqa}{\begin{eqnarray}}
\newcommand{\eeqa}{\end{eqnarray}}
\newcommand{\TeV}{{\rm TeV}}

\def\rhobar{\bar\rho}
\def\etabar{\bar\eta}
\def\lqcd{\Lambda_{\rm QCD}}

\newcommand{\Bbar}{\,\overline{\!B}{}}
\newcommand{\Dbar}{\,\overline{\!D}{}}
\newcommand{\Kbar}{\,\overline{\!K}{}}
\def\B0bar{\Bbar{}^0}
\def\D0bar{\Dbar{}^0}
\def\K0bar{\Kbar{}^0}

\newcommand{\Jpsi}{J\!/\!\psi}
\newcommand{\Gam}[2]{\bar{\Gamma}(#1 \! \to \Jpsi\, #2)}

\title{The Role of Flavor in 2016\hspace*{-1.62cm}
\rotatebox{15}{\rule{44pt}{1.5pt}} 2026}

\ShortTitle{Flavor Constraints on New Physics}

\author{\speaker{Zoltan Ligeti}\\
Lawrence Berkeley National Laboratory,
University of California, Berkeley, CA 94720, USA\\
E-mail: \email{ligeti@berkeley.edu}}

\abstract{This talk explores the role of flavor physics to constrain beyond
standard model phenomena and future prospects, from a theoretical point of
view.  Possible implications of some experimental results in tension with the
standard model are discussed, such as the $4\sigma$ deviation in the $B\to
D^{(*)}\tau\bar\nu$ decay rates.  We use the examples of constraining new
physics contributions to neutral meson mixing and the search for possible
vector-like fermions to illustrate the expected progress over the next decade to
increase the sensitivity to new physics at shorter distance scales.  We also
speculate about the ultimate limitations of (quark) flavor physics probes of new
physics.}

\FullConference{9th International Workshop on the CKM Unitarity Triangle\\
		28  November -- 3 December 2016\\
		Tata Institute for Fundamental Research (TIFR), Mumbai, India}

\begin{document}

\section{Introduction}

I was asked to talk about the role of flavor physics to constraint new physics
(NP).  This is a subject on which there are many different views.  Especially
now, near the end of 2016, the prospects may seem both very bright and somewhat
gloomy.  One the one hand, $K^0-\K0bar$ mixing, and in particular $\Delta m_K$
and $\epsilon_K$ continue to provide some of the best constraints on NP,
unchanged for about 50 years now.  At the same time, the LHC had an amazing year
collecting more than 10 times the data at 13\,TeV than in 2015.  This will yield
a significant increase in the sensitivity to the NP mass scale (at the time of
CKM 2016, or writing this proceedings, we know only the lack of rumors about
discoveries using the 2016 data).  After 2016, however, the next similarly
significant increase in sensitivity will take many years.  In flavor physics,
NA62 took data in 2016 and by the end of this decade $K^+\to \pi^+\nu\bar\nu$
should be measured at the 10\% level if it is near the SM expectation, Belle~II
is rapidly approaching, the LHCb late 2020s upgrade discussion toward 300/fb is
gaining momentum, and there are also bright prospects for significantly improved
sensitivities in electric dipole moment and charged lepton flavor violation
experiments.  So we can look forward to the guaranteed excitement of the
upcoming flood of new data, with many opportunities for groundbreaking
discoveries, definitely testing and understanding the SM much better, as well as
face the uncertainties when NP may be discovered in laboratory experiments.

Going back to the basics, it is important to remember that the SM does not
contain a dark matter candidate, nor can it explain the observed baryon
asymmetry of the Universe.  The solutions of these problems may be connected to
the TeV scale, e.g., the weakly interacting massive particle paradigm and
electroweak baryogenesis, but there are many other options, and there are no
guarantees of accessible discoveries.  We also discovered that neutrinos were
massive; however, the implications of this crucially depend on unraveling
whether neutrino mass terms do or do not violate lepton number (``Majorana vs.\
Dirac").  This leaves the hierarchy puzzle as the clearest connection between
the incompleteness of the SM and our hope to be able to discover NP at the TeV
scale.  If the SM is valid to much higher energy scales than currently probed,
we do not understand why the Higgs particle is so light.  And if there is low
energy supersymmetry, the 126\,GeV Higgs mass seems a bit too high.  So the
situation is confusing and also exciting; or quoting Feynman, ``I think it's
much more interesting to live not knowing than to have answers which might be
wrong."  Not to mention that given the evidence for a nonzero cosmological
constant, one may wonder if even the right questions are being asked about fine
tuning~\cite{sandip}.

In any case, the key question is: What is the scale of NP?  Theoretical
prejudices of the 1990s, when it was often discussed how SUSY cascades would
cause problems to understand LHC signals, now appear wishful thinking.  At the
same time, the evidence for the incompleteness of the SM is compelling, so in
searching for NP, we should leave no stone unturned.  The hierarchy puzzle might
indeed tell us that the NP scale is connected to the electroweak scale, however,
most physicists' measures of fine tuning may be off, and NP could be 1\,--\,2
orders of magnitude heavier than the electroweak scale.  In this case, flavor
physics may be an even more powerful probe of new physics, and its role in
setting future directions even more crucial.  If NP is within the reach of the
LHC, its flavor structure probably has to be fairly similar to that of the SM,
and minimal flavor violation (MFV) is probably a useful notion as a starting
point.  If NP is pushed to the 10\,--\,100\,TeV scale, then the suppressions of
flavor-changing neutral currents (FCNCs) need to be less strong, and MFV becomes
less motivated.  In either case, discovering deviations from the SM in the next
generation of flavor physics experiments is possible, either from LHC-scale NP
with SM-like flavor structure, or from heavier NP with more generic flavor
properties.  Any discovery inconsistent with the standard model would put a
(rough) upper bound on the scale of new physics, which would in turn crucially
impact future directions both in high energy theory and experiment.

\begin{figure}[t]
\centerline{
\includegraphics[width=.5\textwidth, page=1, clip, 
  bb=90 90 540 740]{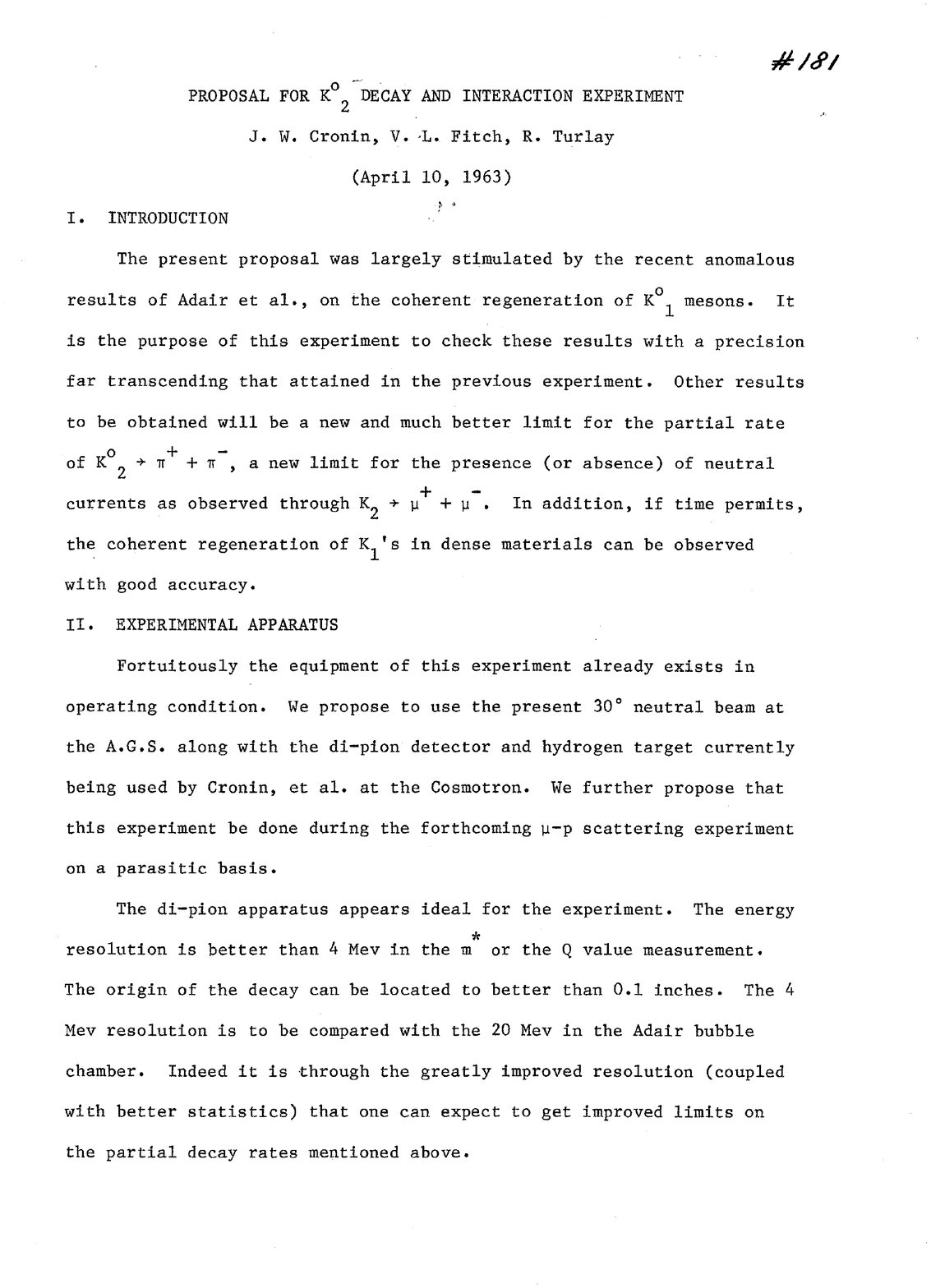} \hfill
\includegraphics[width=.5\textwidth, page=2, clip, 
  bb=90 90 540 740]{CPV_BNLprop}}
\caption{The proposal for the experiment that discovered $CP$ violation in
1964.  (It's legible if you zoom~in.)}
\label{fig:CPVprop}
\end{figure}

Equally importantly, it is hard to anticipate the truly unexpected discoveries,
and one should indeed test all exact and approximate conservation laws as
precisely as possible, especially when the experimental sensitivity can
substantially increase.  After all, the discovery of $CP$ violation itself was
also a surprise, in an experiment whose primary goal was to check an anomalous
kaon regeneration result (read Fig.~\ref{fig:CPVprop}, it is fascinating). Thus,
searches for lepton flavor violation, possible dark sectors in many channels,
are also important parts of future flavor experiments.

In fact, similar surprises did occur at BaBar and Belle --- discovering a suite
of new hadronic states.  One of the most cited BaBar papers is the discovery of
the excited $D_{sJ}^*(2317)$ meson with a mass much below prior
expectations~\cite{Aubert:2003fg}, and the most cited Belle paper is the
discovery of the unexpectedly narrow charmonium-like state
$X(3872)$~\cite{Choi:2003ue} (which will soon surpass in citations the ARGUS
discovery of $B^0 - \B0bar$ oscillation~\cite{Prentice:1987ap} --- so much for
using citations as a measure...).

Section~\ref{sec:status} summarizes the current status of (quark) flavor physics
and reviews some tensions with the SM predictions.  These are some of the most
often discussed topics recently, and they are also interesting because they may
have the best chance to be established as clear deviations from the SM, as more
data is accumulated.  Section~\ref{sec:future} gives some examples of the
expected future progress and improvements in sensitivity to NP, independent of
the current data.  Section~\ref{sec:concl} contains some comments on the
ultimate sensitivity of flavor physics experiments to new physics.

\section{Current status and near future}
\label{sec:status}

\begin{figure}[t]
\centerline{\includegraphics[width=.5\textwidth]{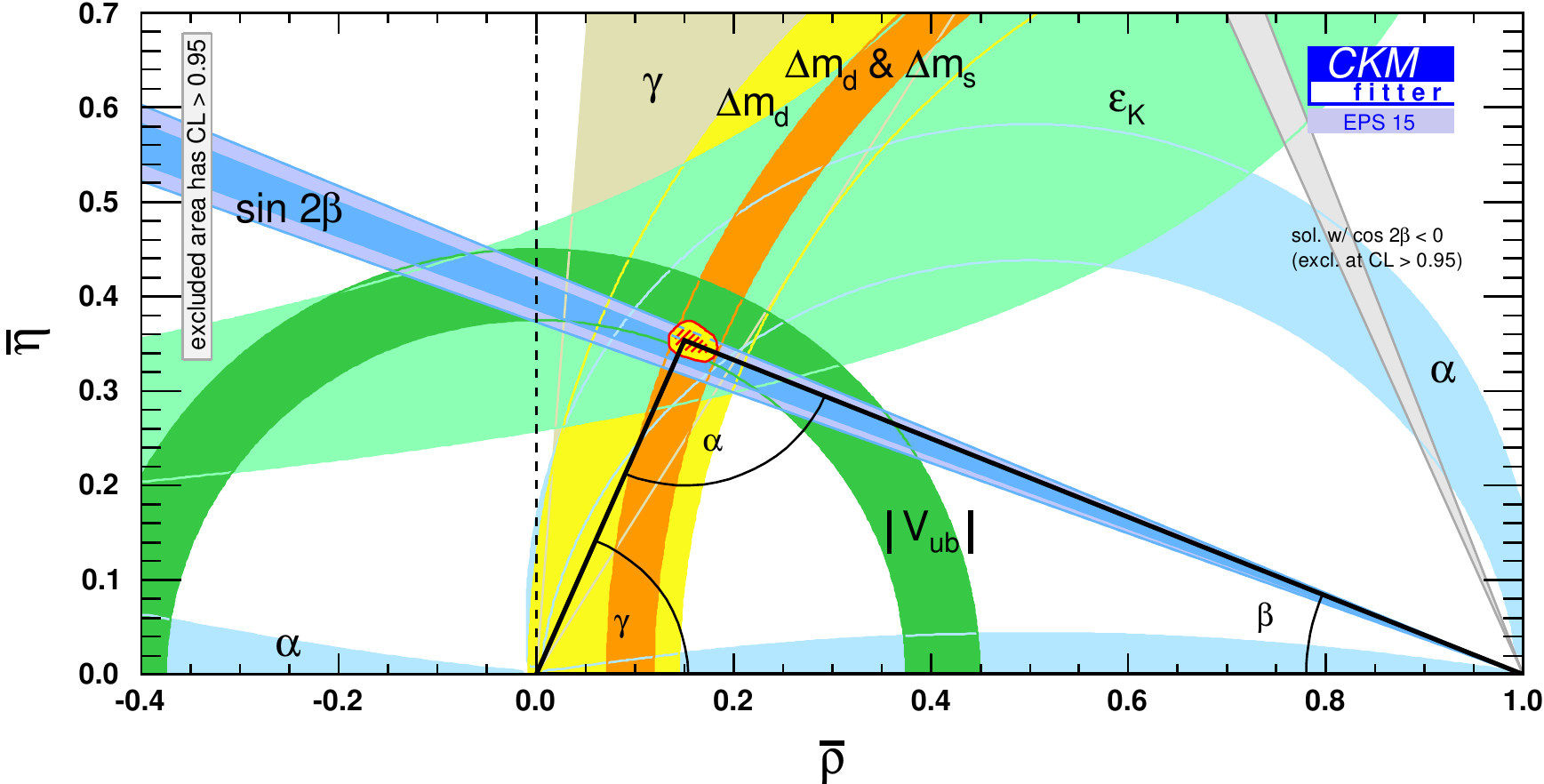} \hfill
\includegraphics[width=.5\textwidth]{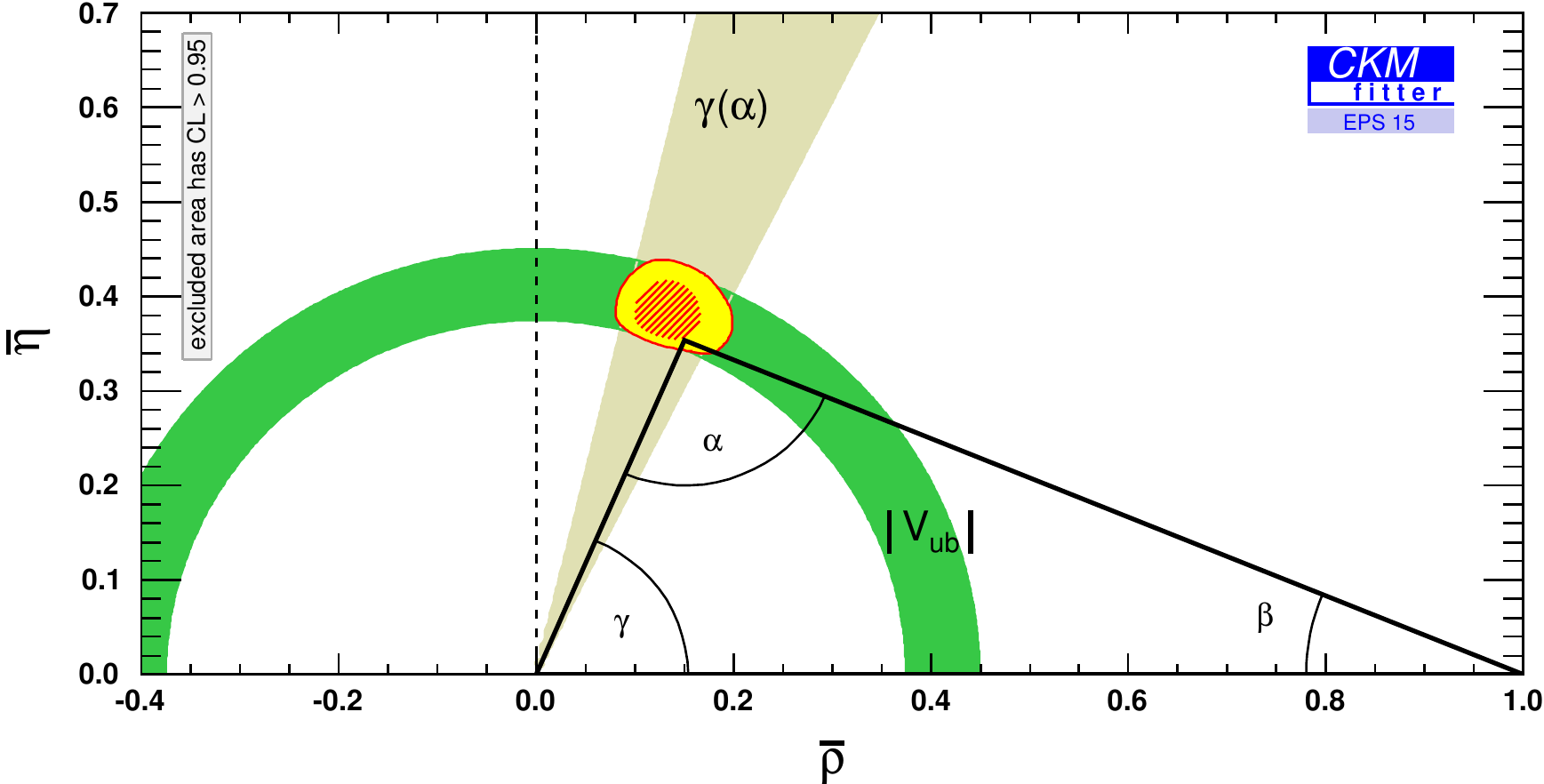}}
\caption{The standard model CKM fit and individual constraints (left); the CKM
fit allowing for new physics in $B^0-\B0bar$ mixing
(right)~\cite{Hocker:2001xe}.  The colored regions show 95\%~CL.}
\label{fig:SMCKMfit}
\end{figure}

A detailed introduction to flavor physics is omitted here, as well as a review
of the determinations of CKM elements; see, e.g., Refs.~\cite{Ligeti:2015kwa,
PDG}. The magnitudes of CKM elements are extracted mainly from semileptonic and
leptonic $K$, $D$, and $B$ decays, and $B_{d,s}$ mixing.  These determine the
sides of the unitarity triangle shown in Fig.~\ref{fig:SMCKMfit} (left), which
is a convenient way to compare many constraints on the SM and visualize the
level of consistency.  Any constraint which renders the area of the unitarity
triangle nonzero, such as nonzero angles (mod $\pi$), has to measure $CP$
violation.  Some of the most important measurements are shown in
Fig.~\ref{fig:SMCKMfit} (left), together with the CKM fit in the SM.  (The
notation $\rhobar,\, \etabar$ instead of $\rho,\, \eta$ corresponds to a small
modification of the Wolfenstein parametrization, to keep unitarity exact.) 
While there is good consistency, that does not address how large new physics
contributions are allowed.  As we see below, in the presence of new physics the
fit becomes less constrained, as shown in Fig.~\ref{fig:SMCKMfit} (right), and
${\cal O}(20\%)$ NP contributions to most FCNC processes, relative to the SM,
are still allowed.

\begin{figure}[t]
\centerline{\includegraphics[width=.5\textwidth]{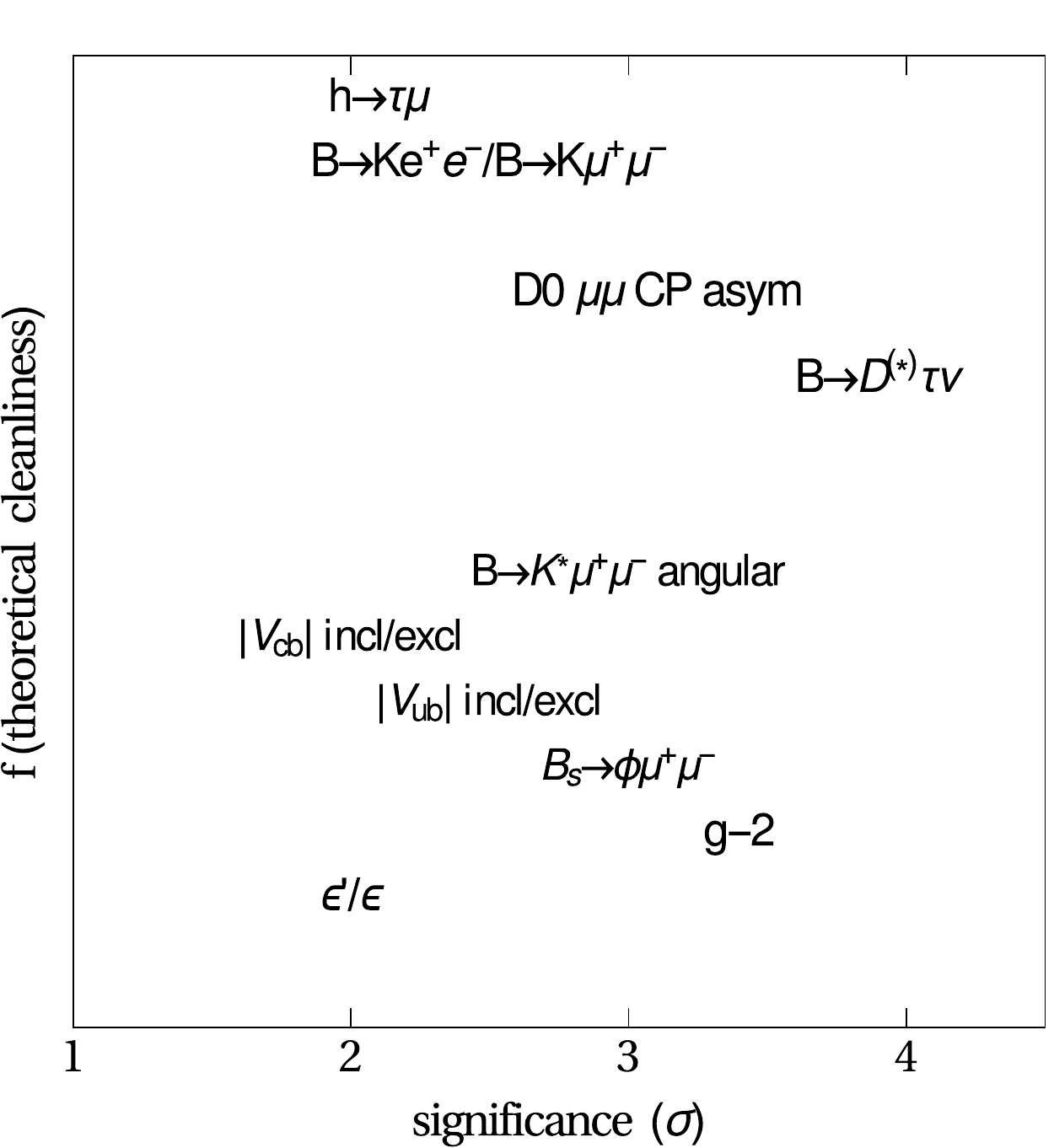}}
\caption{Some recent measurements in tension with the SM.  The horizontal axis
shows the nominal significance.  The vertical axis shows (monotonically, in my
opinion) an undefined function of an ill-defined variable: the theoretical
cleanliness.  That is, the level of plausibility that a really conservative
estimate of the theory uncertainty of each observable may affect the
significance of its deviation from the SM by $1\sigma$.}
\label{fig:cartoon}
\end{figure}

Several measurements show intriguing deviations from the SM predictions.  Some
of those that reach the $2-4\,\sigma$ level are depicted schematically in
Fig.~\ref{fig:cartoon}.  The horizontal axis shows the nominal significance and
the vertical axis relates to the theoretical cleanliness of the SM predictions. 
What I mean is some (monotonic) measure of the plausibility that a conservative
estimate of the theory uncertainty may affect the overall significance by
$1\sigma$.  All of these are frequently discussed, some have triggered hundreds
of papers, and could each be the subjects of entire talks.

\begin{figure}[tb]
\centerline{\small
\begin{tabular}{c|cc}
\hline \hline
&  $R(D)$  &  $R(D^*)$ \\
\hline
World average  &  $0.403 \pm 0.047$  &  $0.310 \pm 0.017$  \\
\hline
SM expectation~\cite{Bernlochner:2017jka}  &  $0.299 \pm 0.005$  &  $0.257 \pm 0.005$  \\
\hline
Belle~II, 50/ab  &  $\pm 0.010$  &  $\pm 0.005$\\
\hline\hline
\end{tabular}\hfill
\raisebox{-64pt}{\includegraphics[width=.45\textwidth, clip, 
  bb=20 20 555 375]{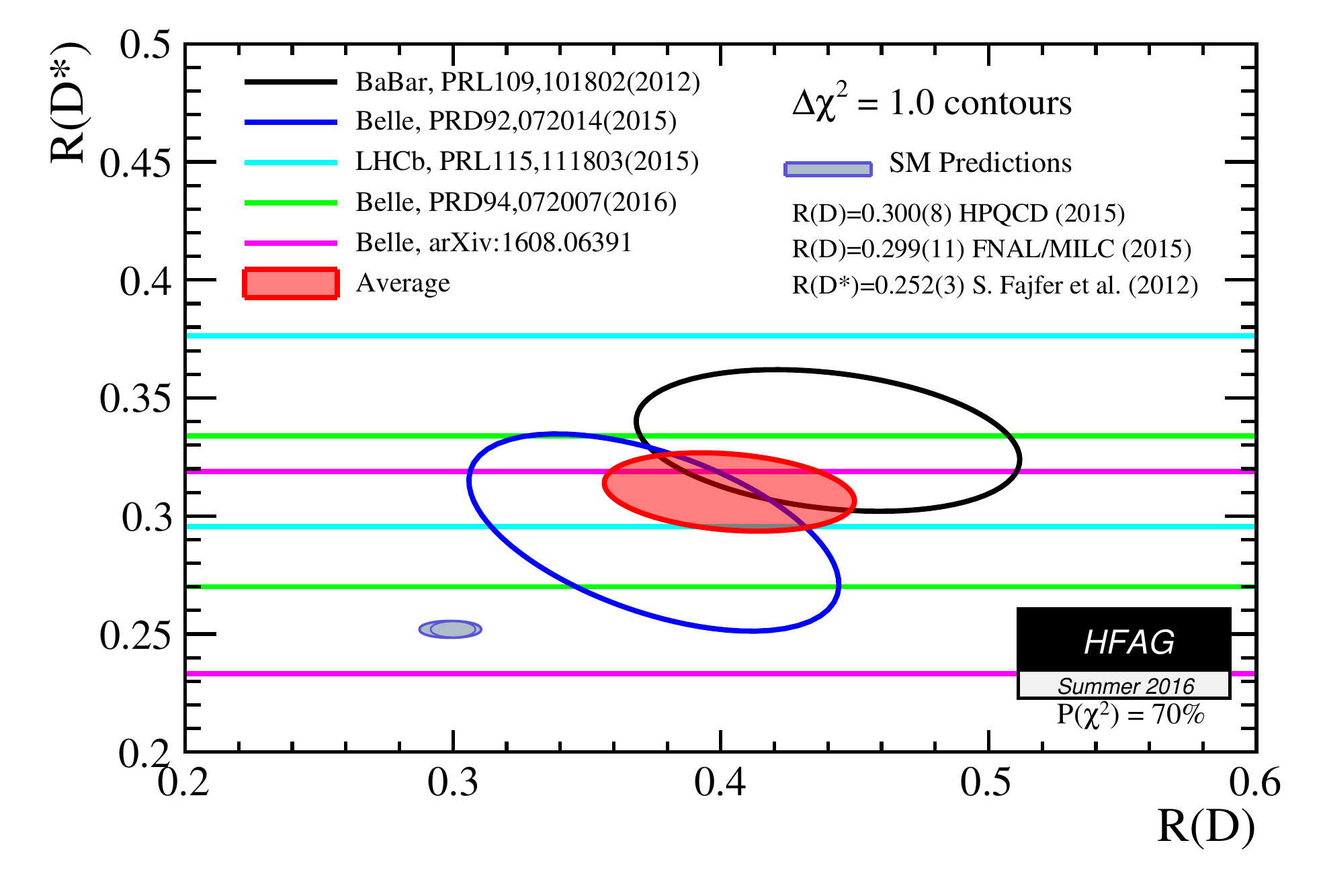}}}
\caption{Left: measurements of $R(D^{(*)})$~\cite{Lees:2012xj, Huschle:2015rga,
Aaij:2015yra, Sato:2016svk, Abdesselam:2016xqt}, their averages~\cite{HFAG}, the
SM predictions~\cite{Bernlochner:2017jka, Lattice:2015rga, Na:2015kha,
Fajfer:2012vx}, and future sensitivity~\cite{Belle2predictions}.  Right: the
measurements, world average (red), and SM prediction.}
\label{fig:RDdata}
\end{figure}

Currently, the $B\to D^{(*)}\tau\bar\nu$ rates, specifically the $R(D^{(*)}) =
\Gamma(B\to D^{(*)}\tau\bar\nu) / \Gamma(B\to D^{(*)}l \bar\nu)$ ratios (where
$l=e,\, \mu$) constitute the most significant discrepancy from the SM in
collider experiments~\cite{Lees:2012xj, Lees:2013udz, Huschle:2015rga,
Aaij:2015yra, Sato:2016svk, Abdesselam:2016xqt} (aside from neutrino masses). 
The effect is at the $4\sigma$ level~\cite{HFAG}.  Figure~\ref{fig:RDdata} shows
the current data, the SM expectations, as well as the expected Belle~II
sensitivity.  These measurements show good consistency with one another.  The
theory is also on solid footing, since heavy quark symmetry suppresses model
independently the hadronic physics needed for the SM prediction, most of which
is constrained by the measured $B\to D^{(*)}l \bar\nu$ decay
distributions.

It is somewhat surprising to find so large deviations from the SM in processes
which occur at tree level.  The central values of the current world averages
would imply that there has to be new physics at or below the TeV scale.  Some
scenarios are excluded by LHC Run 1 bounds already, and more will soon be
constrained by the Run 2 data.  To fit the current central values, mediators
with leptoquark or $W'$ quantum numbers are preferred, compared to scalars. 
Leptoquarks are favored if one requires the NP to be minimally flavor violating
(MFV), which helps explain the absence of other flavor signals and suppress
direct production of the new particles at the LHC from partons abundant in
protons~\cite{Freytsis:2015qca}.  Currently the ``simplest" models that fit the
data modify the SM four-fermion operator (after Fierzing), and then the $\tau$
polarization is not affected, in agreement with its first
measurement~\cite{Abdesselam:2016xqt}.  There are even viable scenarios in which
$B\to D^{(*)}\tau\bar\nu$ are SM-like, but $B\to D^{(*)} l \bar\nu$ are
suppressed by interference between NP and the SM~\cite{future}.  

There are many further measurements that may help to clarify this anomaly.  The
$B\to D^{(*)}\tau\bar\nu$ rates seem to exceed~\cite{Freytsis:2015qca} the LEP
measurements of the inclusive $b\to X \tau \bar\nu$ rate~\cite{PDG}, and the
inclusive $B\to X_c\tau\nu$ rate~\cite{Ligeti:2014kia} has not yet been
measured.  The $B\to D^{**}\tau\bar\nu$ rates will also give complementary
information~\cite{Bernlochner:2016bci}.  The equality of the $e$ and $\mu$ rates
are not well constrained, and the currently allowed
differences~\cite{Aubert:2008yv, Dungel:2010uk} open up (or keep open) many
model building options~\cite{Greljo:2015mma}.  In many scenarios, bounds on
$b\to s\nu\bar\nu$ processes are very important~\cite{Freytsis:2015qca,
Fortes:2015jaa}.  A lot will be learned, hopefully soon, from the comparison of
LHCb and Belle~II data with fully differential theory
predictions~\cite{Ligeti:2016npd}.  If a deviation from the SM is established,
it will strongly motivate to measure all possible semitauonic modes, both in
$b\to c$ and $b\to u$ transitions~\cite{Bernlochner:2015mya, Hamer:2015jsa}.

\begin{figure}[t]
\centerline{\includegraphics[width=.47\textwidth]{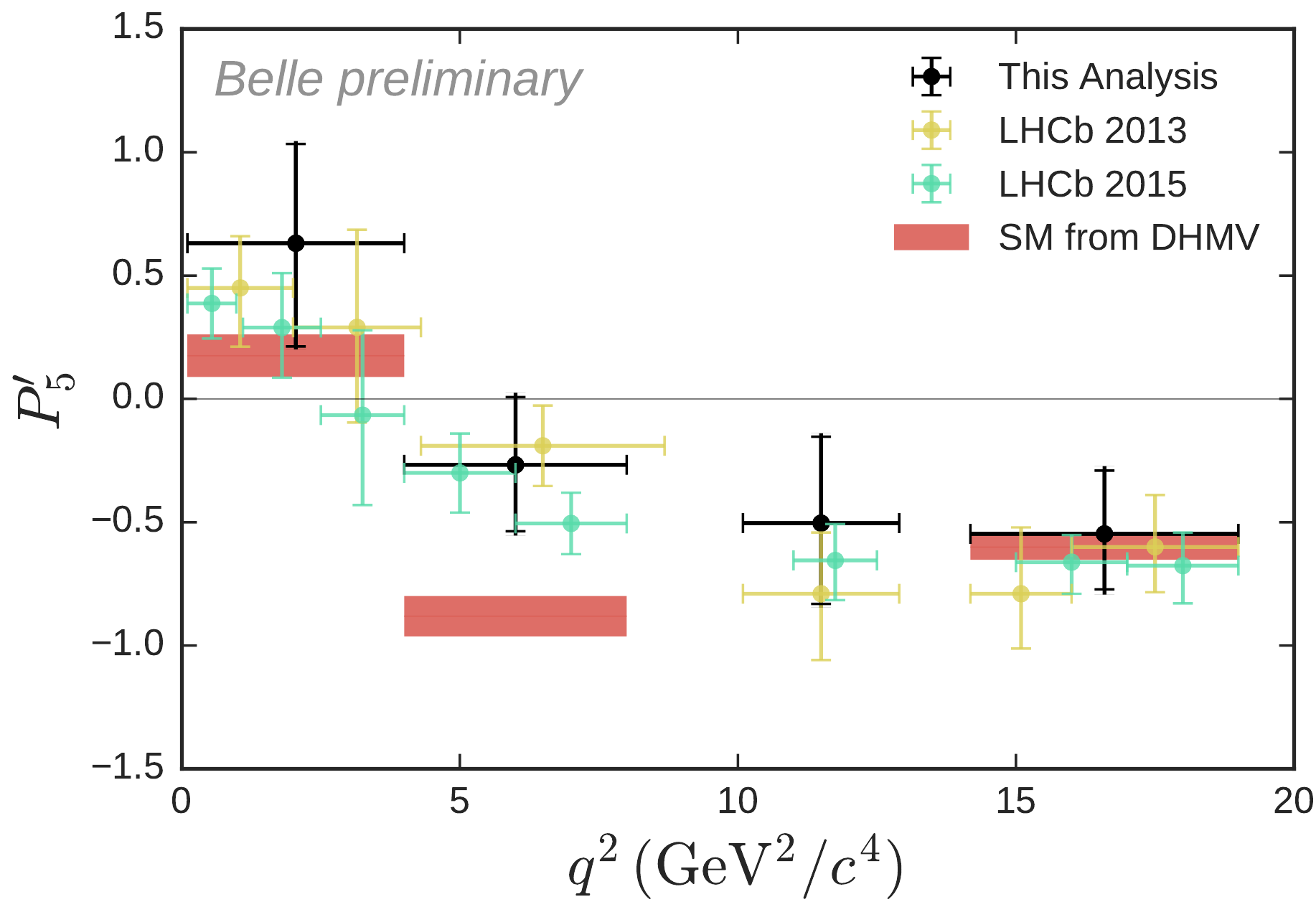}\hfill
\includegraphics[width=.515\textwidth]{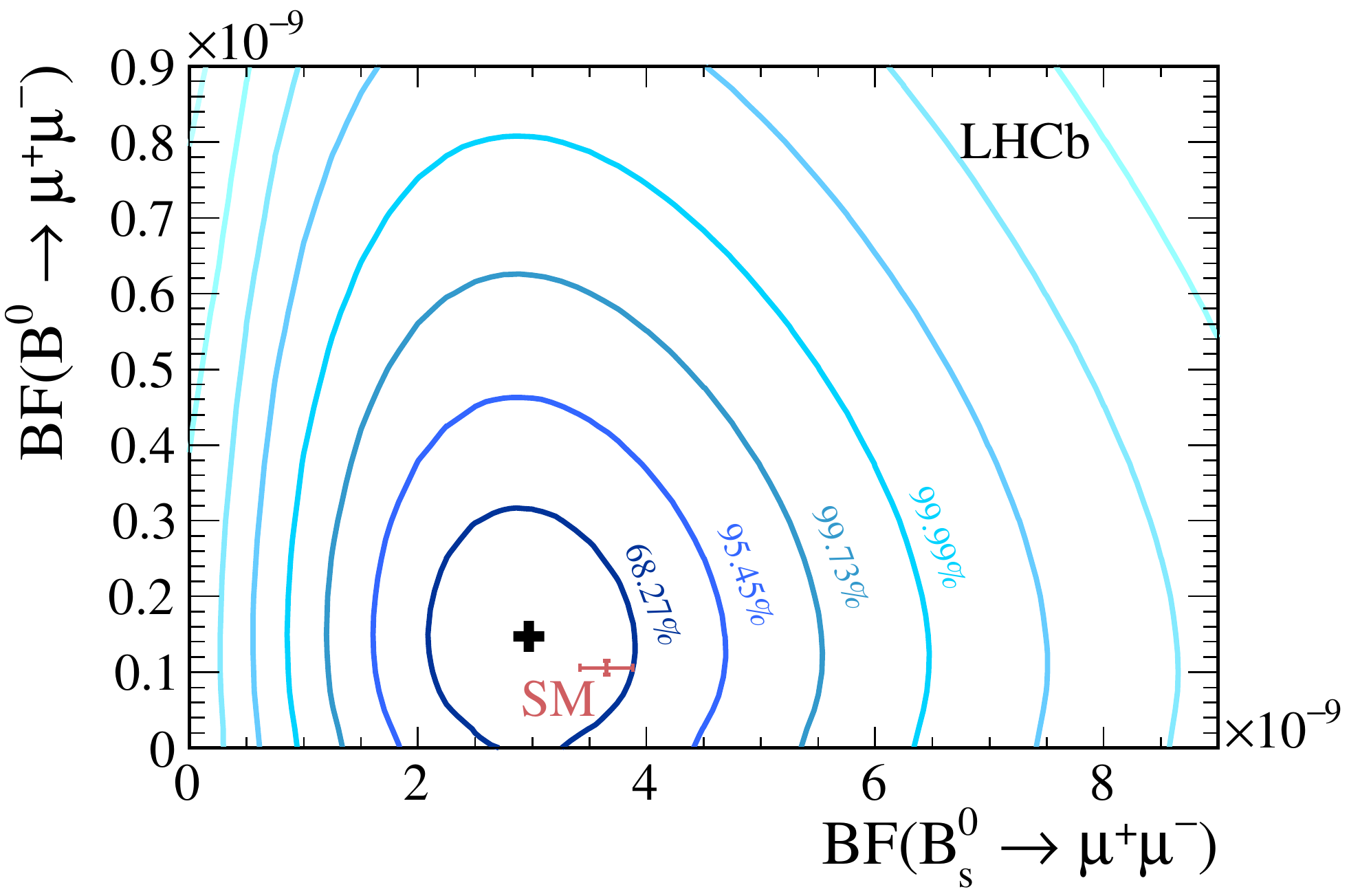}}
\caption{Left: The LHCb~\cite{Aaij:2015oid} and Belle~\cite{Abdesselam:2016llu}
measurements of $P'_5$ in $B\to K^*\mu^+\mu^-$. 
Right: The $B_{s,d} \to\mu^+\mu^-$ result from LHCb~\cite{Aaij:2017vad}
(average with CMS and ATLAS measurements~\cite{CMS:2014xfa, Aaboud:2016ire} is
not available).}
\label{fig:KsBsmumu}
\end{figure}

Another measurement which has drawn immense attention is the ``$P'_5$ anomaly"
in a $B\to K^*\mu^+\mu^-$ angular distribution (see, e.g.,
Refs.~\cite{Descotes-Genon:2013wba, Altmannshofer:2014rta}), measured first at
LHCb~\cite{Aaij:2015oid} and then at Belle~\cite{Abdesselam:2016llu}.  The
measurements are shown in the left plot in Fig.~\ref{fig:KsBsmumu}, together
with a SM prediction~\cite{Descotes-Genon:2014uoa}.  These ``optimized
observables" are based on the SCET factorization theorem for semileptonic $B$
decay form factors~\cite{Bauer:2002aj, Beneke:2003pa}, and constructing
combinations from which the ``nonfactorizable" (``soft") contributions cancel. 
(These are nonperturbative functions of $q^2$, which obey symmetry
relations~\cite{Charles:1998dr}; additional terms are either power suppressed or
contain an explicit $\alpha_s$ factor.)  The magnitudes of the correction terms,
that is one's ability to calculate the form factor ratios at small $q^2$
reliably, is debated~\cite{Jager:2014rwa} (and not well constrained by data
yet).  The tension between theory and the data is intriguing.  Some of the
simplest new physics explanations are $Z'$-like models, with nonuniversal flavor
couplings.  One may be concerned that the best fit is a new contribution to the
operator $O_9 = e^2 (\bar s \gamma_\mu P_L b) (\bar\ell \gamma^\mu \ell)$ in the
effective Hamiltonian, the same term which would be modified if theoretical
control over the $c\bar c$ loop contributions were worse than expected.  (This
was also emphasized recently in Ref.~\cite{Ciuchini:2015qxb}.)  There are many
possible connections to the $\sim 2.5\sigma$ anomaly in $\Gamma(B\to K e^+e^-)
\neq \Gamma(B\to K \mu^+\mu^-)$ as well~\cite{Gudrun}.

For these observables, too, I trust that with improved measurements and theory,
the source of the currently seen effects will be understood.  With more data,
one can test  the $q^2$ (in)dependence of the extracted Wilson coefficients. In
the large $q^2$ (small recoil) region one can make model independent predictions
both for exclusive~\cite{Bobeth:2012vn} inclusive~\cite{Ligeti:2007sn} $b\to s
l^+l^-$ mediated decays, which is complementary to the small $q^2$ region, and
has different theory uncertainties.

If new physics is at play in these processes, it is likely to impact $B\to
\mu^+\mu^-$, too.  The very recent LHCb measurement~\cite{Aaij:2017vad} shown in
the right plot in Fig.~\ref{fig:KsBsmumu} is consistent with the SM, and no
longer hints at an enhancement of $B_d\to\mu^+\mu^-$~\cite{CMS:2014xfa}. 
Measuring a rate at the $3\times 10^{-9}$ level is impressive, and future
refinements are high priority.  The nonperturbative input in this case is just
$f_B$, which is under good control in lattice QCD.

\begin{figure}[t]
\centerline{\includegraphics[width=.38\textwidth]{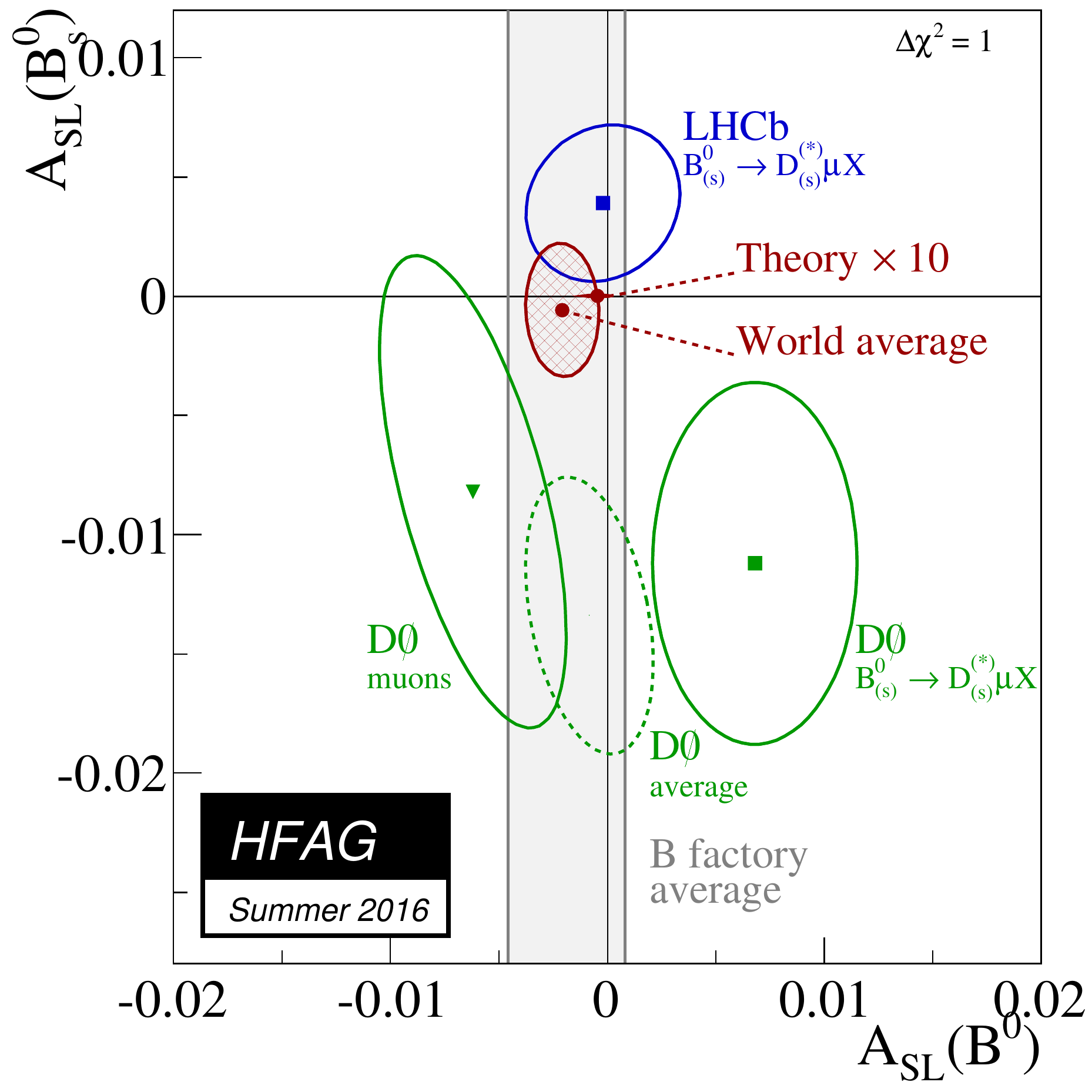} \hfil\hfil
\raisebox{4pt}{\includegraphics[width=.57\textwidth]{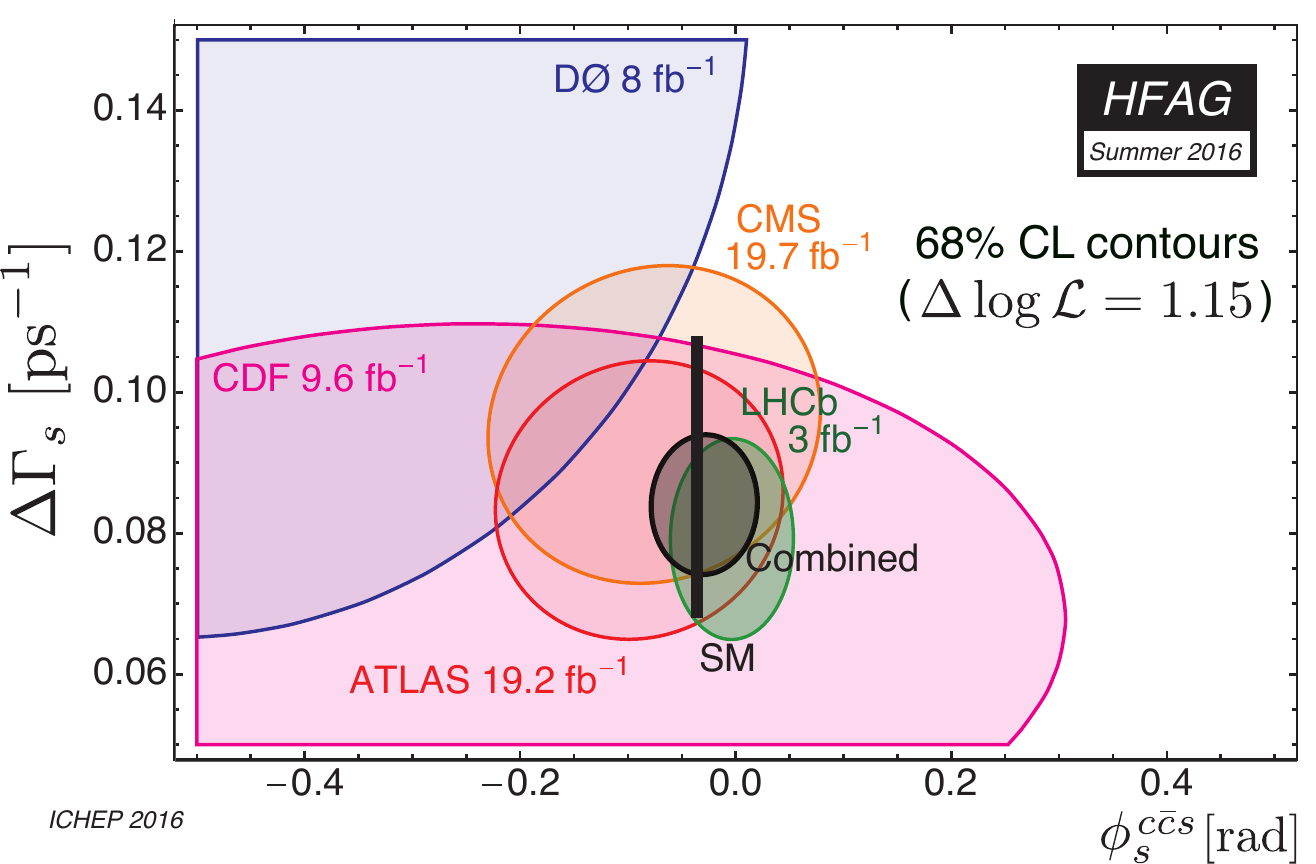}}}
\caption{Left: bounds on $CP$ violation in $B_{d,s}$ mixing, $a_{\rm
SL}^{d,s}$. The vertical and horizontal bands show the
averages of the separate $B_d$ and $B_s$ measurements, respectively, and the
yellow ellipse is the D\O\ measurement.   Right: measurements of $\phi_s \equiv
-2\beta_s$ showing good consistency with the SM.}
\label{fig:cpvmix}
\end{figure}

Another deviation from the SM expectations, which is theoretically very clean,
and has been $3-4\,\sigma$, is the D\O\ measurement of the like-sign dimuon
charge asymmetry in semileptonic decays of $b$ hadrons, $(N_{\mu^+\mu^+} -
N_{\mu^-\mu^-}) / (N_{\mu^+\mu^+} + N_{\mu^-\mu^-})$~\cite{Abazov:2011yk}, shown
in the left plot in Fig.~\ref{fig:cpvmix}.  A nonzero signal could come from a
linear combination of $CP$ violation in $B_s$ and $B_d$ mixing, $a_{\rm
SL}^{d,s}$ (see, e.g., Ref.~\cite{Ligeti:2010ia}), and the SM prediction is well
below the current sensitivity.  Separate measurements of $a_{\rm SL}^d$ and
$a_{\rm SL}^s$ from BaBar, Belle, and D\O\ are consistent in with the SM, and 
the recent LHCb update with 3/fb, $a_{\rm SL}^s = (0.39 \pm
0.33)\%$~\cite{Aaij:2016yze}, starts to be in tension with the D\O\ anomaly.  If
there is new physics in $CP$ violation in $B_s$ mixing, then one may also expect
to see a deviation from the SM in the time-dependent $CP$ asymmetry in $B_s\to
\Jpsi\phi$ and in related modes.  Recent LHC measurements, however, are
consistent with the SM, as shown in the right plot in Fig.~\ref{fig:cpvmix}. 
Most importantly, the theory uncertainties are well below the experimental
sensitivity in the coming years, so a lot can be learned from more precise
measurements.

It has long been known that kaon $CP$ violation is sensitive to some of the
highest energy scales.  For the $\epsilon$ parameter, the SM is in good
agreement with the data, and the NP contribution is constrained to be $\,\lsim
30\,\%$ of that of the SM~\cite{Ligeti:2016qpi}.  Calculating the SM prediction
for direct $CP$ violation, the $\epsilon'$ parameter, has been a multi-decade
challenge, and progress is being made~\cite{Bai:2015nea}.  Results with several
lattice spacings are needed to decide if NP is present. My views of the
theoretical status of the measurements shown in Fig.~\ref{fig:cartoon}, but not
discussed here, are explained in Ref.~\cite{Ligeti:2016riq}.

These experimental hints of possible deviations from the SM are fantastic for
several reasons.  Unambiguous evidence for NP would obviously be the start of a
new era, and would also provide a rough upper bound on the scale of NP, even if
it is not seen directly at ATLAS \& CMS.  It is also useful to have experimental
results challenge theory, since unexpected signals motivate both model building
and revisiting the SM predictions.  This was the case with the Tevatron anomaly
in the $t\bar t$ forward-backward asymmetry, $A_{\rm FB}^{t\bar t}$, which
disappeared due to refinements of the experimental results (the SM predictions
also improved~\cite{Czakon:2014xsa}).  Concerning the recent $3\,\sigma$ hint
for direct $CP$ violation in the difference of $CP$ asymmetries in $D\to K^+K^-$
and $D\to \pi^+\pi^-$, $\Delta A_{CP} = A_{K^+K^-} - A_{\pi^+\pi^-}$, I doubt
the initial measurement near $1\%$ could be attributed to the
SM~\cite{Isidori:2011qw}.  The central value of the world average has decreased
since 2012, as has the significance of the hint for $\Delta A_{CP} \neq 0$.  We
probably still do not know how large $\Delta A_{CP}$ the SM could generate. 
However, exploring it taught us, for example, about how much (or how little) the
quark and squark mixing matrices can differ and squark masses (don't) need to be
degenerate~\cite{Gedalia:2012pi, Mahbubani:2012qq} in alignment
models~\cite{Nir:1993mx}.

\begin{figure}[t]
\centerline{\includegraphics[width=.52\textwidth]{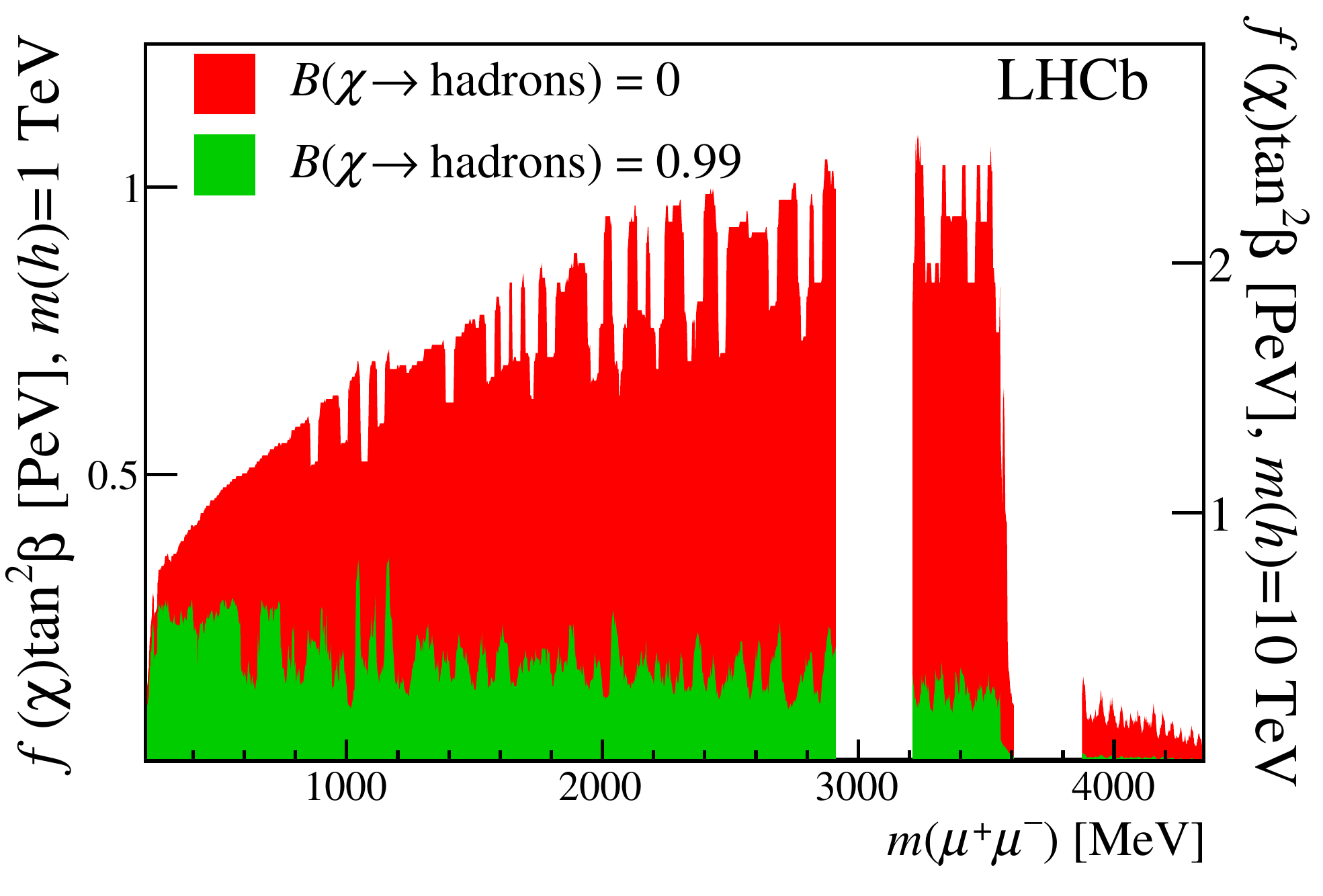} \hfill
\includegraphics[width=.45\textwidth]{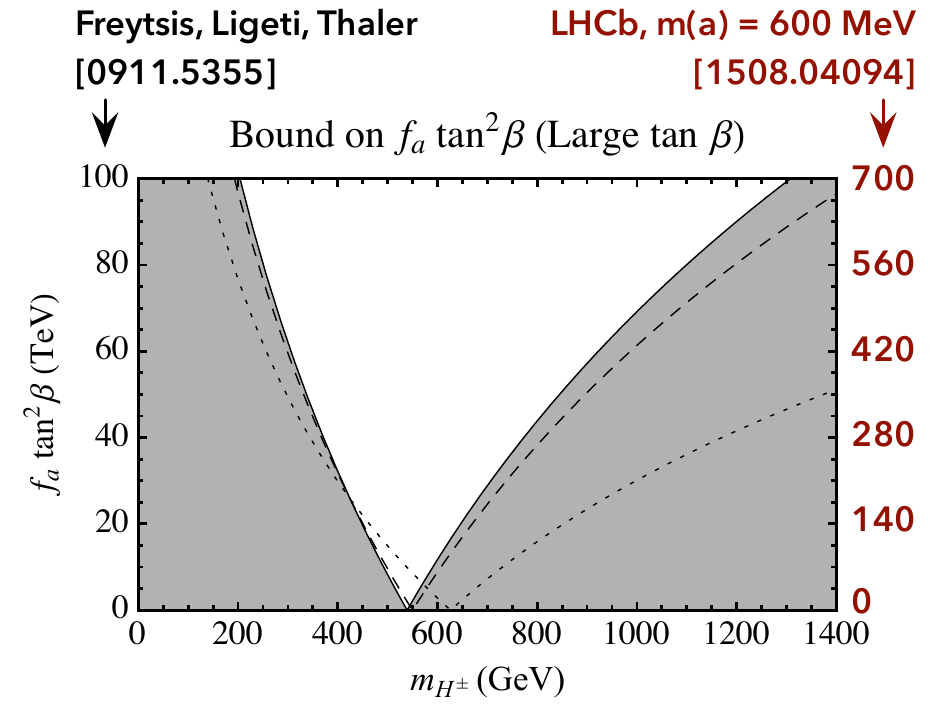}}
\caption{Left: LHCb bounds on $f_\chi^2\, \tan^2\beta$ as a function of
$m_{\mu+\mu^-}$~\cite{Aaij:2015tna} in the model~\cite{Freytsis:2009ct}.
Right: The bound as a function of $m_{H^\pm}$ in the same model; the right axis
shows a nearly order of magnitude improvement.}
\label{fig:axionsearch}
\end{figure}

A measurement in which no anomaly is seen, but there is a nearly order of
magnitude increase in mass-scale sensitivity due to a recent LHCb
analysis~\cite{Aaij:2015tna}, is the search for an axion-like particle, coupling
to SM fermions as $(m_\psi/f_a)\, a\, \bar\psi\gamma_5\psi$.  Such models are
also interesting, because they may have highly suppressed spin-independent
direct detection cross sections~\cite{Freytsis:2010ne}. The left plot in
Fig.~\ref{fig:axionsearch} shows the 95\% CL lower bound on $f_\chi^2\,
\tan^2\beta$ in the model of Ref.~\cite{Freytsis:2009ct}, from the absence of a
narrow $\mu^+\mu^-$ peak in $B\to K^*\chi\,(\chi\to \mu^+\mu^-)$ as a function
of $m_{\mu^+\mu^-}$.  The bound is shown for $m_{H^\pm} = 1\,\TeV$ and two
values of the hadronic branching fraction of the axion-like particle.  The right
plot shows the bound on the same quantity as a function of $m_{H^\pm}$ ($f_a$
in~\cite{Freytsis:2009ct} is $f_\chi$ in~\cite{Aaij:2015tna}).  The left
vertical axis is the bound estimated in 2009~\cite{Freytsis:2009ct} from BaBar
\& Belle data with only a few bins, and the right vertical axis shows the LHCb
bound~\cite{Aaij:2015tna}.  The dashed (dotted) curve shows the bound for
$\tan\beta = 3$ ($\tan\beta = 1$).  In this model, for any value of $\tan\beta$,
the NP contribution vanishes due to a cancellation for a certain value~of
$m_{H^\pm}$.  There are promising proposals to utilize the upcoming huge LHCb
data sets for related dark photon searches as well~\cite{Ilten:2016tkc,
Ilten:2015hya}

\section{Future increases in new physics scales probed}
\label{sec:future}

I would like to talk about three topics briefly in this part: (i) the future
theory uncertainty of the measurement on $\sin2\beta$ from $B\to \Jpsi K_S$;
(ii) the future sensitivity to NP in mixing of neutral mesons; (iii) sensitivity
of flavor physics experiments to very heavy vector-like fermions.

\subsection{What is the ultimate theory uncertainty of $\mathbf{\sin2\beta}$?}

The theoretical uncertainty of the SM predictions for the time dependent $CP$
asymmetries in the ``gold-plated" modes $B\to \Jpsi K_S$ and $B_s\to \Jpsi \phi$
are of great importance.  They arise from contributions to the decay amplitude
proportional to $V_{ub}V_{us}^*$ instead of the dominant $V_{cb}V_{cs}^*$ terms.
I refer to this as $V_{ub}$ contamination, instead of the often used penguin
pollution phrase (which is less correct and less clear).  This effect did not
matter in practice in the past, but it will be important for interpreting the
full LHCb and Belle II data sets.   During the BaBar/Belle era, the experimental
precision was an order of magnitude above the nominal magnitude of the
theoretical uncertainty, $\lambda^2 (\alpha_s/\pi) \sim 0.004$.  So even a
factor of few enhancement of the latter did not matter.\footnote{Until about
1997 this was often estimated as $\lambda^2 (\alpha_s/4\pi)$. Omitting the
factor 4 anticipates some enhancement of the penguin matrix element, observed in
charmless $B$ decays~\cite{Godang:1997we} but not yet well constrained in decays
to charmonia.  Calculable ${\cal O}(10^{-3})$ effects arise from $CP$ violation
in $K$ and $B$ mixing, and the $\Gamma_{B_L} - \Gamma_{B_H}$ width
difference~\cite{Grossman:2002bu}.}  A number of approaches have been developed,
using a combination of diagrammatic and flavor symmetry arguments with various
assumptions~\cite{Jung:2012mp, Frings:2015eva}.  (I hasten to add a triviality:
there is no relation based only on $SU(3)$ flavor symmetry between final states
which are entirely in different representations; e.g., $\phi$ is an $SU(3)$
singlet and $\rho$ \& $K^*$ are members of an octet.)  The experimental tests
performed so far do not indicate big enhancements of the theory uncertainties.

The question that really matters in my opinion is not what it takes to set
plausible upper bounds on the $V_{ub}$ contamination, when the measurements
agree with the SM, but what it would take to convince the community that NP is
observed at LHCb and Belle~II, especially if no NP is seen by ATLAS and CMS. 
Therefore, one cannot overemphasize the importance of starting from rigorous
theoretical foundations, with well defined expansion parameter(s).  

\begin{figure}[t]
\centerline{\includegraphics[width=.45\textwidth]{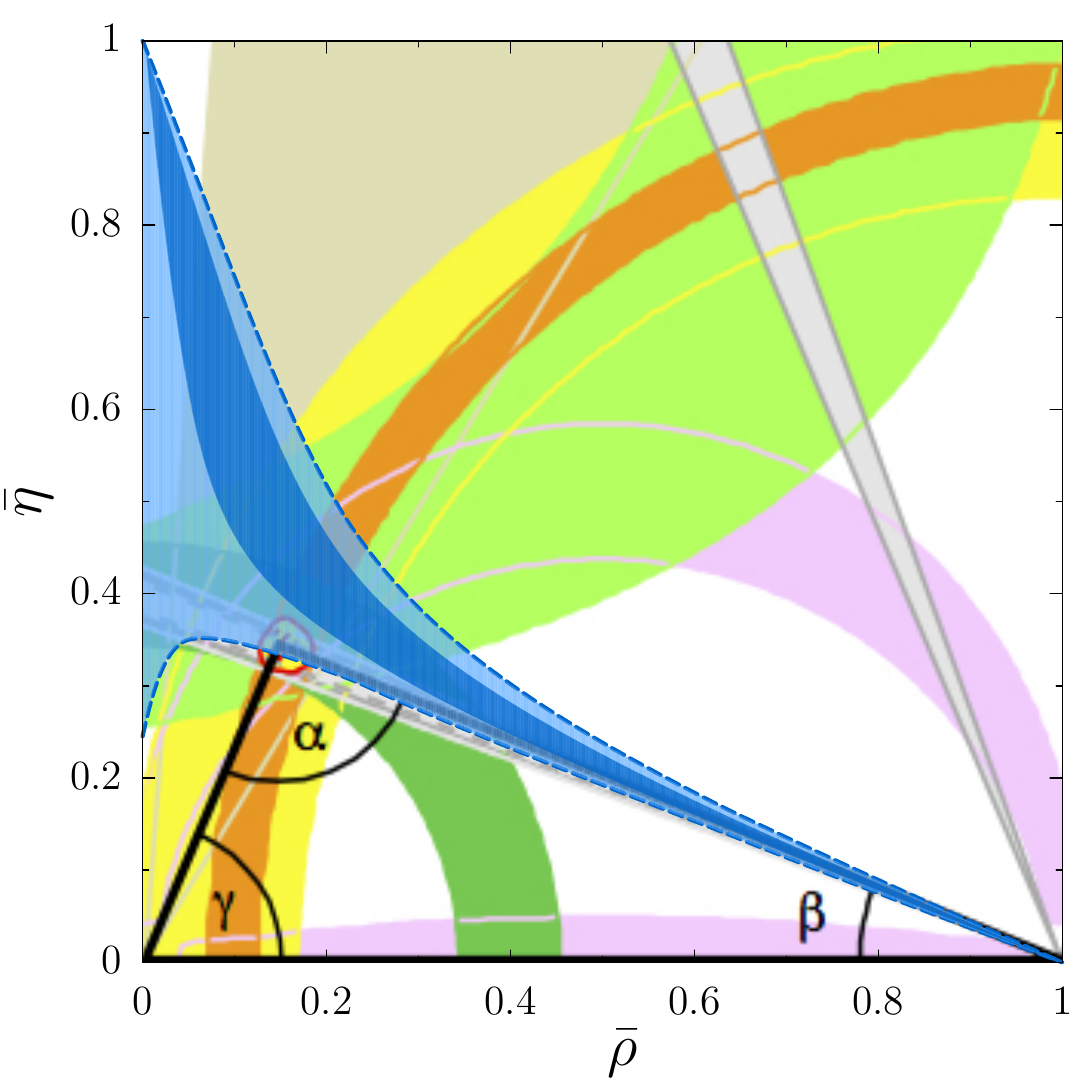}}
\caption{The dark (light) blue region shows the $1\sigma$ ($2\sigma$) constraint
in the $\rhobar - \etabar$ plane from
Eq.~(\protect\ref{beauty})~\cite{Ligeti:2015yma}.}
\label{fig:s2bfuture}
\end{figure}

A relation based only on $SU(3)$ flavor symmetry, which cancels the $V_{ub}$
contamination in $\sin2\beta$ against other observables in the $SU(3)$ limit,
reads~\cite{Ligeti:2015yma}
\beq\label{beauty}
\sin 2\beta = \frac{S_{K_S} - \lambda^2 S_{\pi^0}
  - 2(\Delta_K + \lambda^2 \Delta_\pi) \tan\gamma\, \cos2\beta}{1 +
  \lambda^2}\,.
\eeq
Here $S_h$ ($h=K,\,\pi$) is the usual coefficient of the $\sin(\Delta m\, t)$
term in the time-dependent $CP$ asymmetry~\cite{PDG} in $B\to \Jpsi\, h^0$,
$\lambda\simeq 0.225$ is the Wolfenstein parameter,
\beq
\Delta_h = \frac{\Gam{B_d}{h^0} - \Gam{B^+}{h^+}}
  {\Gam{B_d}{h^0} + \Gam{B^+}{h^+}} \,,
\eeq
and $\bar\Gamma$ denotes the $CP$ averaged rates.  Using Eq.~(\ref{beauty}), it
is possible to replace the $V_{ub}$ contamination in the $\sin2\beta \simeq
S_{K_S}$ relation with isospin breaking, which could be smaller than the
possibly enhanced $V_{ub}$ contamination one wants to constrain.  It also
provides redundancy, replacing one theory uncertainty with a different one.  For
the $V_{cb}V_{cs}^*$ terms in the effective Hamiltonian, $\Delta_{K,\pi}$
violate isospin, but the $V_{ub}V_{us}^*$ terms generate nonzero $\Delta_h$ even
in the isospin limit.  The resulting constraint on the $\rhobar - \etabar$ plane
is shown in Fig.~\ref{fig:s2bfuture}~\cite{Ligeti:2007sn}.

Measuring all terms in Eq.~(\ref{beauty}) is not straightforward.  Many of the
current measurements of $\Delta_h$ and the production asymmetry of $B^+B^-$ vs.\
$B^0\B0bar$ in $\Upsilon(4S)$ decay, $f_{+-}/f_{00}$, are circular (the
measurements of either assume that the other asymmetry
vanishes)~\cite{Jung:2015yma}, so the slight tension in Fig.~\ref{fig:s2bfuture}
should be interpreted with caution.  To disentangle $\Delta_h$ from the
production asymmetry, more precise measurements of the latter are needed.  One
option may be to utilize that isospin violation in inclusive semileptonic decay
is suppressed by $\lqcd^2/m_b^2$~\cite{Jung:2015yma}.  (Similar suppression of
$SU(3)$ symmetry breaking in inclusive $B$ decays by $\lqcd^2/m_b^2$ is the
basis for a theoretically clean prediction for the ratio $\Gamma(B\to
X_s\ell^+\ell^-) / \Gamma(B\to X_u\ell \bar\nu)$ at large
$q^2$~\cite{Ligeti:2007sn}.)

It is an open question how well it will be possible to ultimately constrain
(convincingly) the size of $V_{ub}$ contamination in the measurements of
$\sin2\beta$ and $2\beta_s (\equiv -\phi_s)$.  Given that $SU(3)$ flavor
symmetry has been used to analyze $B$ decays for decades, and previously unknown
$SU(3)$ relations can be discovered in 2015, makes me optimistic that a lot more
progress can be achieved.

\subsection{New physics in SM loop processes}

\begin{figure}[t]
\centerline{
\includegraphics[height=6cm,clip,bb=15 15 550 520]{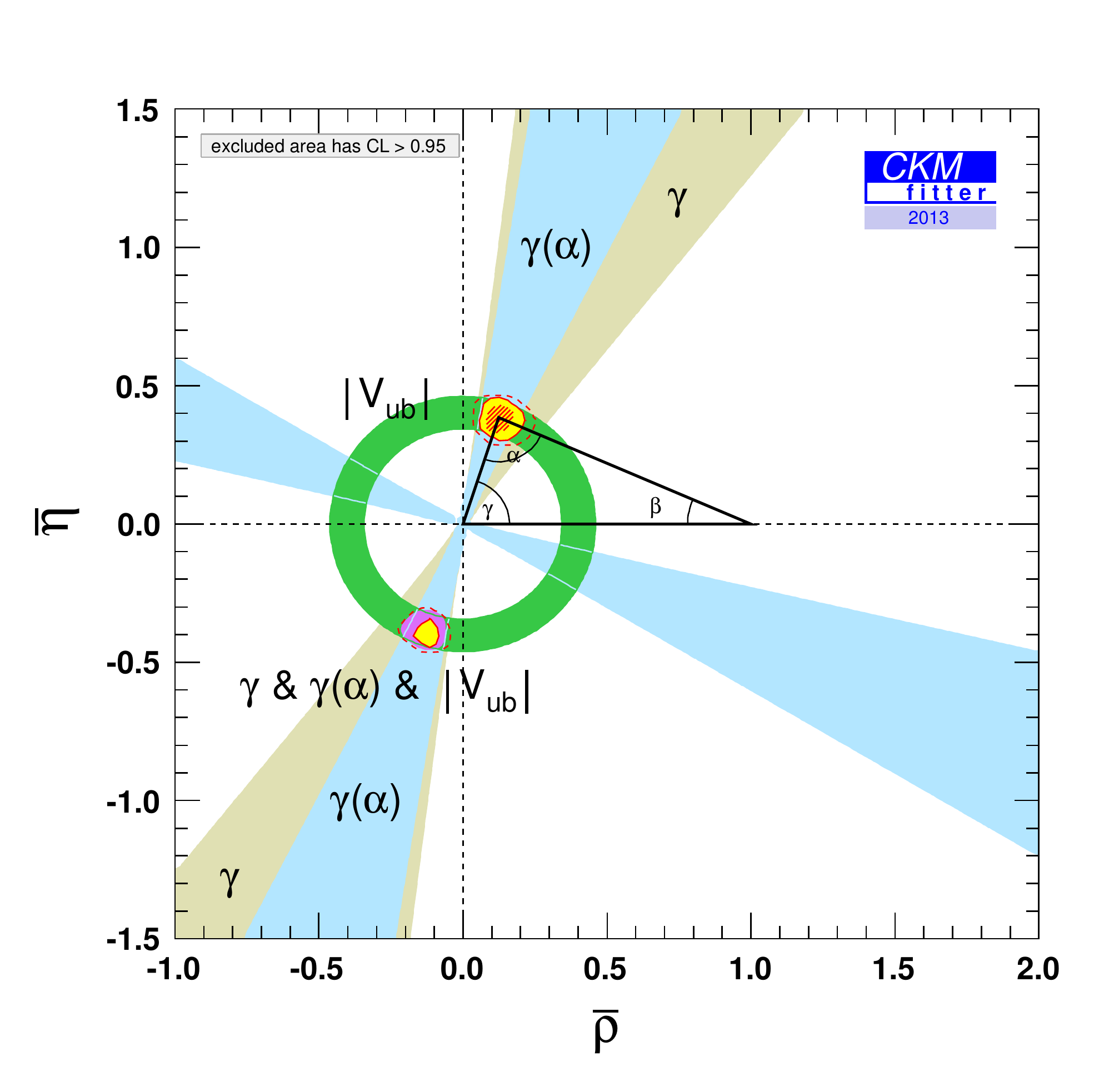}
\hfil\hfil
\includegraphics[height=6cm,clip,bb=105 20 670 550]{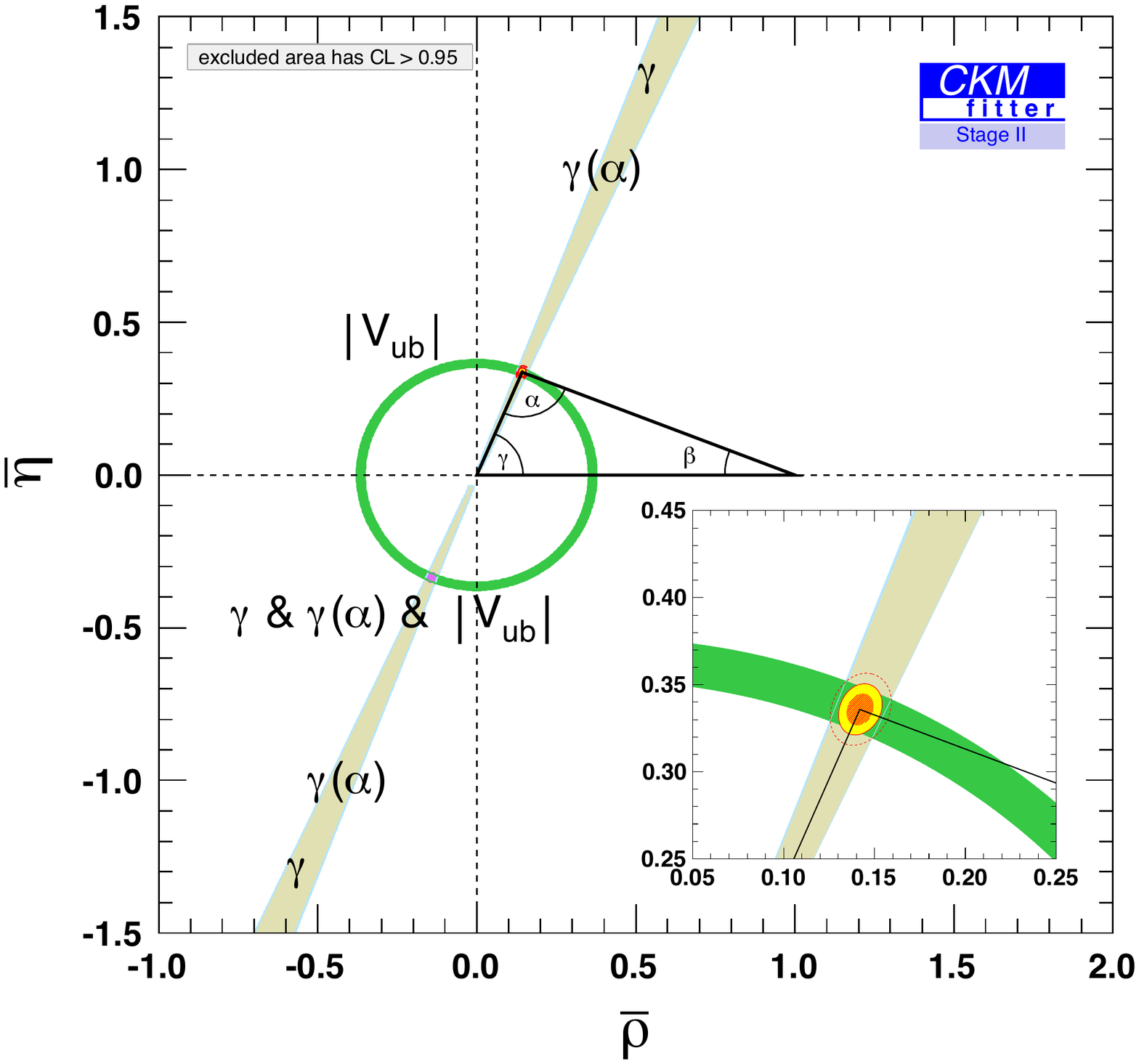}}
\caption{Constraints on $\rhobar - \etabar$, allowing NP in the $B_{d,s}$ mixing
amplitudes (left) and the expectation using 50\,ab$^{-1}$ Belle~II and
50\,fb$^{-1}$ LHCb data (right)~\cite{Charles:2013aka}.  Colored regions show
95\% CL, as in Fig.~\protect\ref{fig:SMCKMfit}.}
\label{fig:NPrhoeta}
\end{figure}

Although the SM CKM fit in Fig.~\ref{fig:SMCKMfit} shows impressive and
nontrivial consistency, the implications of the level of agreement are often
overstated.  Allowing new physics contributions, there are a larger number of
parameters related to $CP$ and flavor violation, and the fits become less
constraining.  This is shown in Fig.~\ref{fig:NPrhoeta}, which shows the
determination of the unitarity triangle from tree-dominated decays only, which
are unlikely to be affected by new physics.  The plot on the left shows the
current fit results, while the constraints in the plot on the right is expected
to be achievable with 50\,ab$^{-1}$ Belle~II and 50\,fb$^{-1}$ LHCb
data~\cite{Charles:2013aka}.  The allowed region in the left plot is indeed
significantly larger than in Fig.~\ref{fig:SMCKMfit}.

\begin{figure}[tb]
\centerline{
\includegraphics[height=6cm,clip,bb=15 15 490 449]{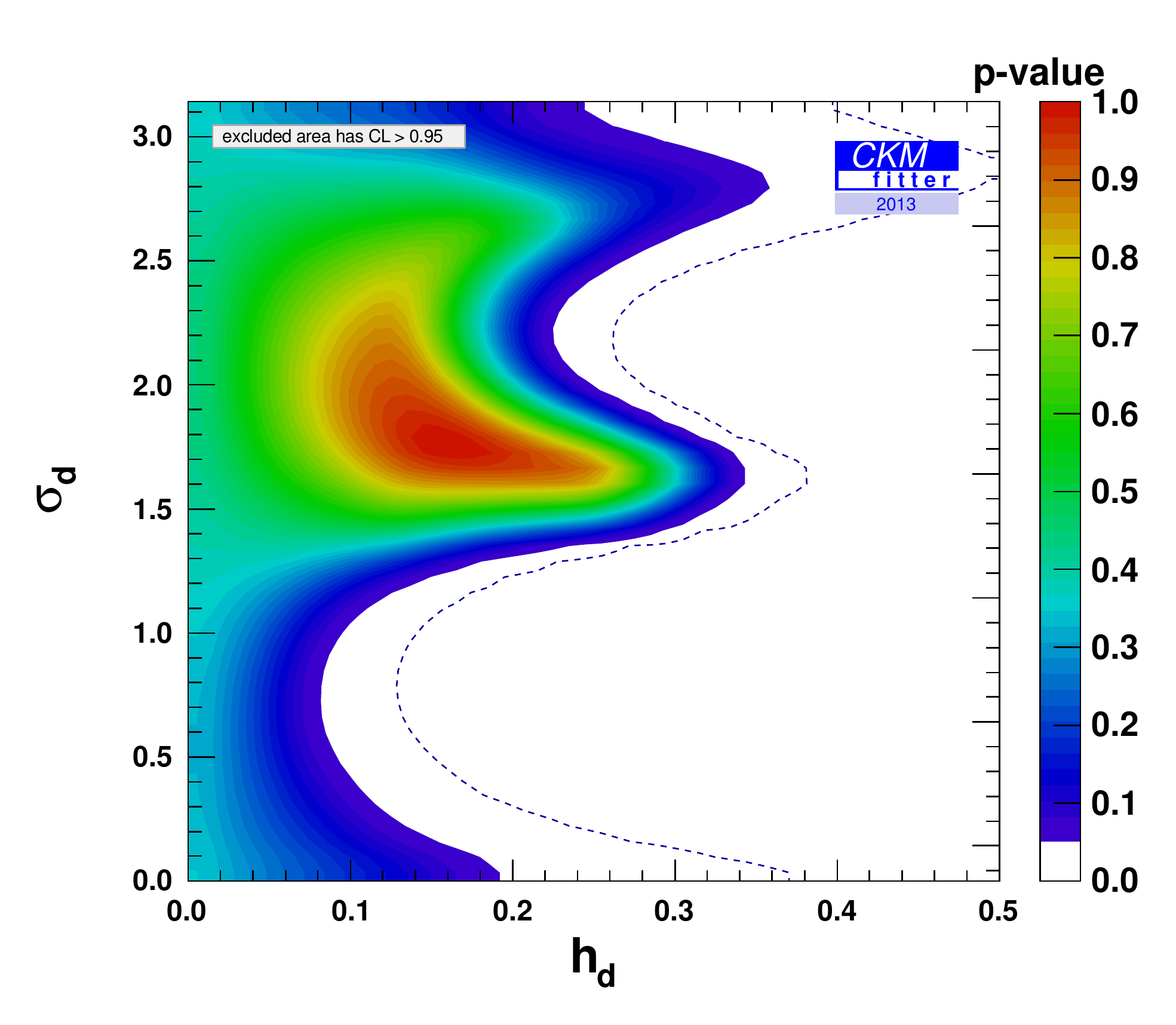}
\hfil\hfil
\includegraphics[height=6cm,clip,bb=15 15 550 449]{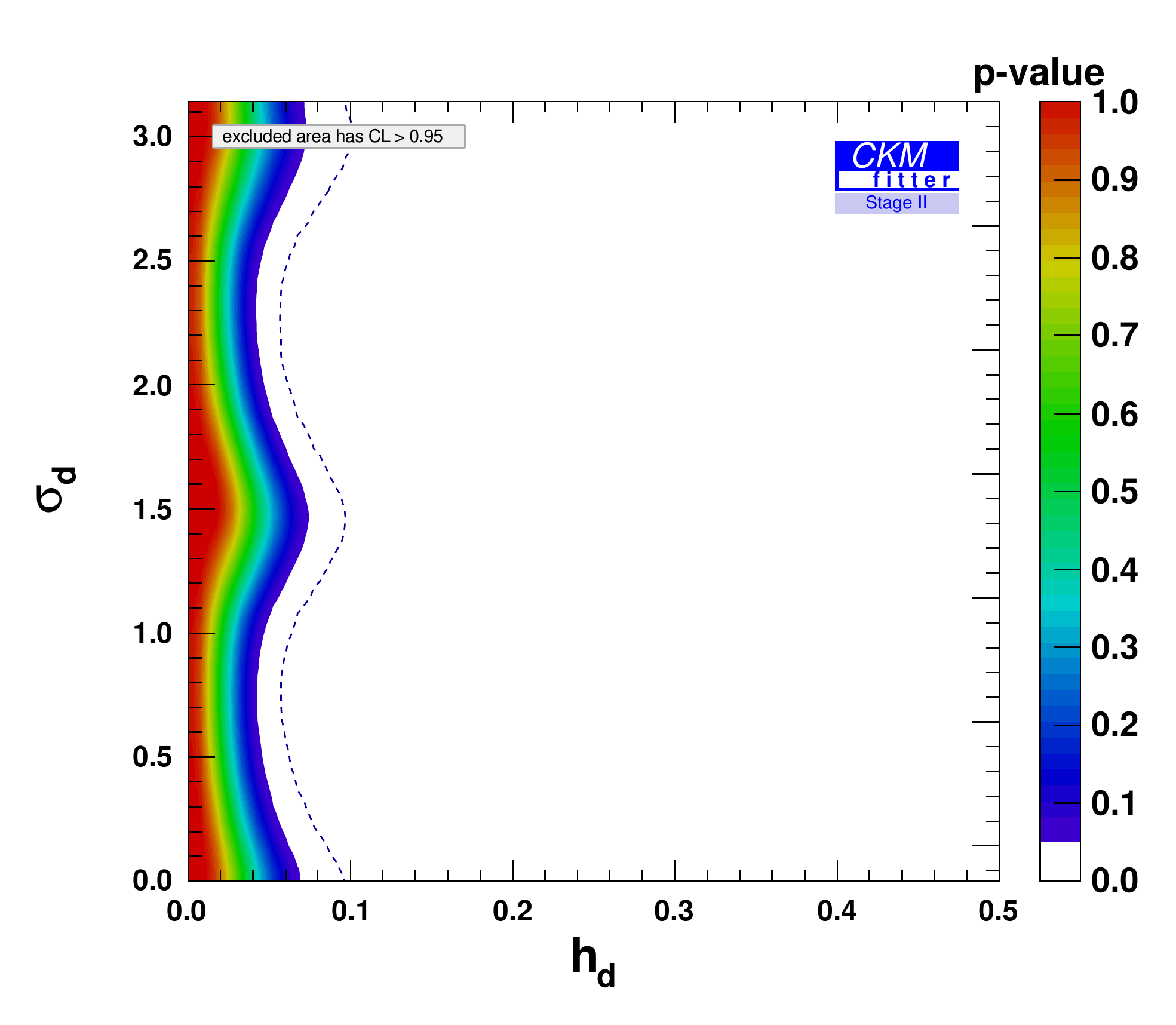}}
\caption{Constraints on the $h_d - \sigma_d$ parameters (left) and those
estimated to be achievable using 50\,ab$^{-1}$ Belle~II and 50\,fb$^{-1}$ LHCb
data (right)~\cite{Charles:2013aka}.  Colored regions show $2\sigma$ limits with
the colors indicating CL as shown, while the dashed curves show $3\sigma$
limits.}
\label{fig:NPBdmix}
\end{figure}

\begin{figure}[tb]
\centerline{
\includegraphics[height=6cm,clip,bb=15 15 490 449]{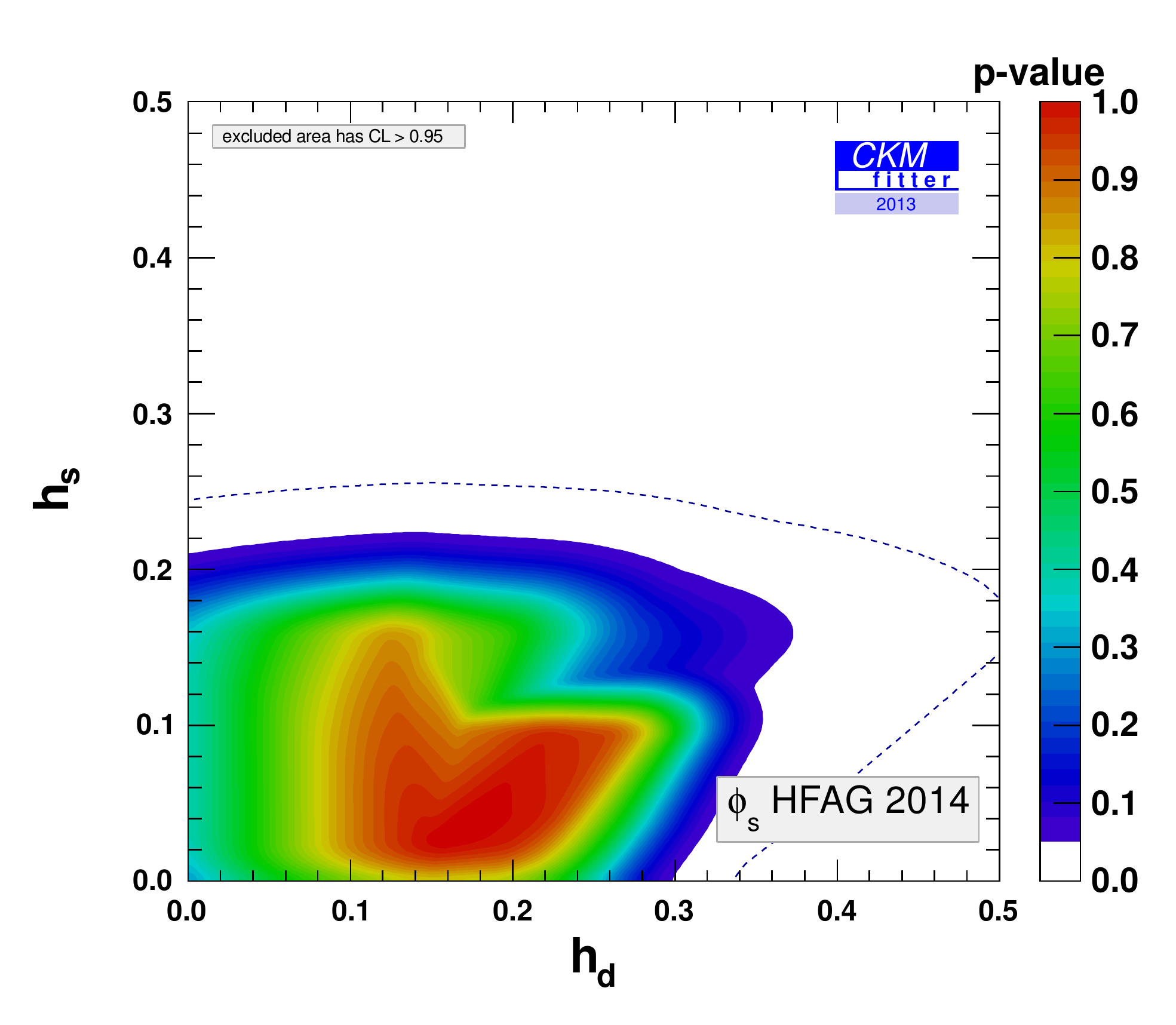}
\hfil\hfil
\includegraphics[height=6cm,clip,bb=15 15 550 449]{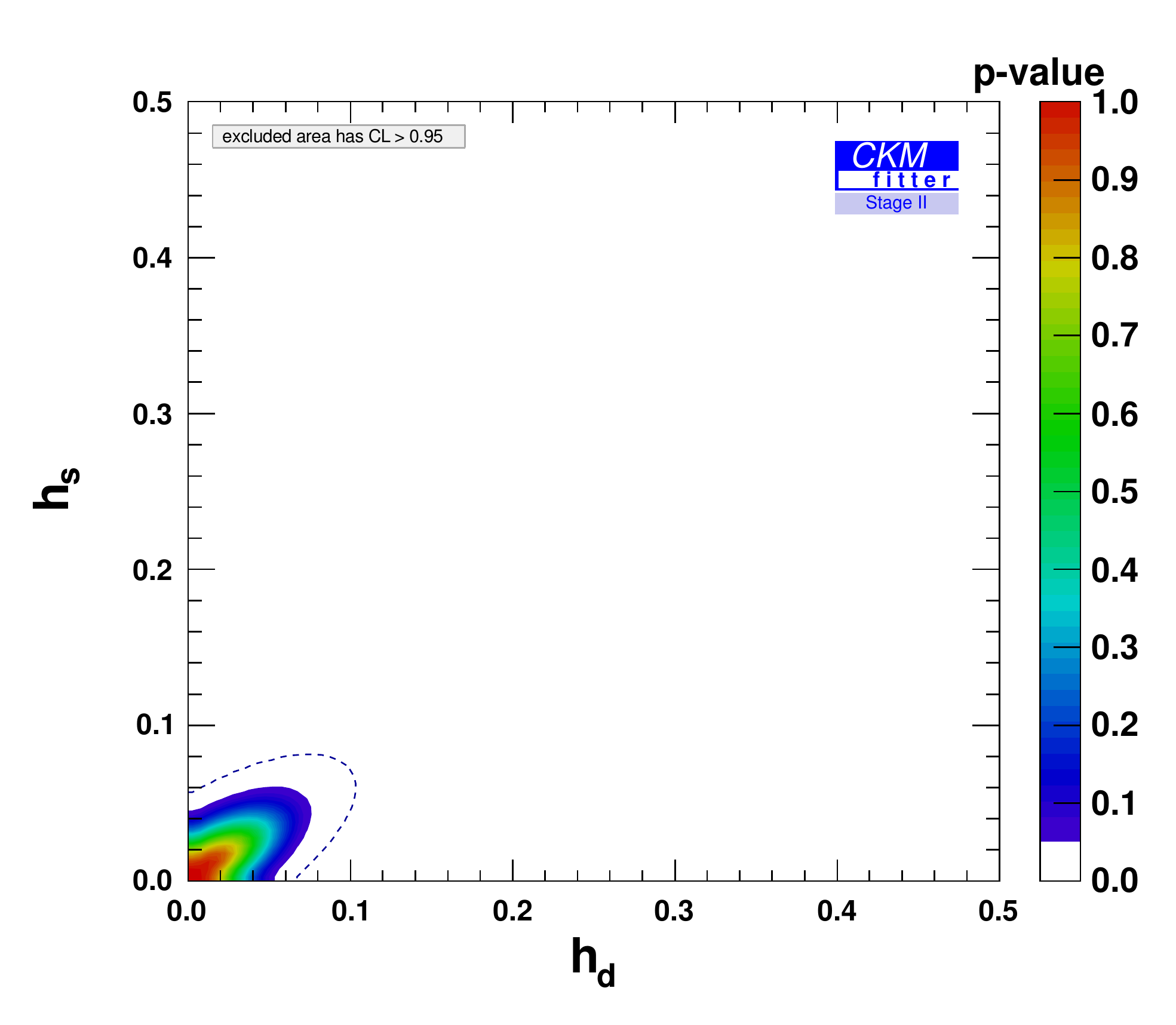}}
\caption{Constraints on the $h_d - h_s$ parameters now (left plot) and those
estimated with 50\,ab$^{-1}$ Belle~II and 50\,fb$^{-1}$ LHCb data (right
plot)~\cite{Charles:2013aka}.  The notation is the same as in
Fig.~\protect\ref{fig:NPBdmix}.}
\label{fig:NPBdBsmix}
\end{figure}

It has been known for decades that the mixing of neutral mesons is particularly
sensitive to new physics, and probe some of the highest scales.  In a large
class of models, NP has a negligible impact on tree-level SM transitions (e.g.,
the measurements of $\gamma$, $|V_{ub}|$, and $|V_{cb}|$), and the $3\times 3$
CKM matrix remains unitary.  As a simple example, consider possible NP
contributions to $B$ and $B_s$ meson mixing, which can be parametrized as
\beq\label{M12param}
M_{12} = M_{12}^{\rm SM} (1 + h_q\, e^{2i\sigma_q})\,, \qquad q=d,s\,.
\eeq
The constraints on $h_d$ and $\sigma_d$ in the $B_d$ mixing are shown in
Fig.~\ref{fig:NPBdmix}, and the constraint in the $h_d - h_s$ plane is shown in 
Fig.~\ref{fig:NPBdBsmix}.  Both plots show the current constraints (left) and
those expected to be achievable with 50\,ab$^{-1}$ Belle~II and 50\,fb$^{-1}$
LHCb data (right)~\cite{Charles:2013aka}.  Figure~\ref{fig:NPBdmix} shows that
in the future the bounds on the ``MFV-like regions", where NP flavor is aligned
with the SM ($2\theta_d \simeq 0$ mod $\pi$), will be comparable to  generic
values of the NP phase, unlike in the past.  Figure~\ref{fig:NPBdBsmix} shows
that the bounds on NP in $B_s$ mixing, which were significantly weaker than
those in the $B_d$ sector until recent LHCb measurements, are now comparable,
and will comparably improve in the future.

As an example, if NP modifies the SM operator describing $B_q$ mixing
by adding to it a term
\beq\label{SMmix}
\frac{C_{q}^2}{\Lambda^2}\, (\bar b_{L}\gamma^{\mu}q_{L})^2\,,
\eeq
then one finds
\beq
h_q \simeq \frac{|C_{q}|^2}{|V_{tb}^*\, V_{tq}|^2} 
  \left(\frac{4.5\, \TeV}{\Lambda}\right)^2\,.
\eeq
We can then translate the plotted bounds to the scale of new physics probed. 
The summary of expected sensitivities are shown in Table~\ref{scaletable}. The
sensitivities even with SM-like loop- and CKM-suppressed coefficients are
comparable to the scales probed by the LHC in the next decade.

\begin{table}[t]\tabcolsep 14pt
\renewcommand{\arraystretch}{1.2}
\centerline{
\begin{tabular}{c|c|c|c}
\hline\hline
\multirow{2}{*}{Couplings}  &  NP loop  &  
  \multicolumn{2}{c}{Scales (TeV) probed by}\\
&  order  &  $B_d$ mixing  &  $B_s$ mixing   \\
\hline
$|C_{q}| = |V_{tb}V_{tq}^*|$ & tree level &  17  &  19 \\
\cdashline{2-4}
(CKM-like)  &  one loop  &  1.4  &  1.5  \\
\hline
$|C_{q}| = 1$  &  tree level  &  $2\times 10^3$  &  $5\times 10^2$  \\
\cdashline{2-4}
(no hierarchy)  &  one loop &  $2\times 10^2$  &  40 \\
\hline\hline
\end{tabular}}
\caption{The scale of the $B_{d,s}$ mixing operators in
Eq.~(\protect\ref{SMmix}) probed, with 50\,ab$^{-1}$ Belle~II and 50\,fb$^{-1}$
LHCb data~\cite{Charles:2013aka}.  The differences due to CKM-like hierarchy of
couplings and/or loop suppression is shown.}
\label{scaletable}
\end{table}

In $K^0$\,--\,$\K0bar$ mixing the simplest analog of Eq.~(\ref{M12param}) is to
parametrize NP via an additive term to the so-called $tt$ contribution in the
SM, $M^{K,tt}_{12} = M^{K,tt}_{12} (1 + h_K\, e^{2i\sigma_K})$.  The reason is
the short distance nature of NP and the fact that in many NP models the largest
contribution to $M^K_{12}$ arise via effects involving the third generation. 
Substantial progress would require lattice QCD to constrain the long distance
contribution to $M_{12}^K$ at the percent level~\cite{Charles:2013aka}.

There are also strong constraints on NP from $D^0$\,--\,$\D0bar$ mixing.  Since
the observed mixing parameters are probably dominated by long distance
physics~\cite{Falk:2001hx}, it is hard to improve the bound from simply
demanding the NP contribution to be below the measured values of the mixing
parameters.

\subsection{Sensitivity to vector-like fermions}

Another illustration of the expected progress with well quantifiable increases
in mass scale sensitivity,  in both quark and lepton flavor experiments, is to
consider extensions of the SM involving vector-like fermions, which can Yukawa
couple to the SM~\cite{Ishiwata:2015cga}.  These fermions can have masses $M$
much greater than the weak scale, since they have a mass term even in the
absence of electroweak symmetry breaking.  These models are a class of simple
extensions of the SM, which do not worsen the hierarchy puzzle.  There are 11
renormalizable models~\cite{Ishiwata:2015cga} which add to the SM vector-like
fermions in a single (complex) representation of the gauge group that can Yukawa
couple to the SM fermions through the Higgs field (4 to leptons, 7 to quarks).

The precise definitions of the $\lambda_i$ Yukawa couplings depend on the
models, as do the forms of the Lagrangians.  For example, what was labeled
Model~V in Ref.~\cite{Ishiwata:2015cga} contains vector-like fermions, $D$, with
the same quantum numbers as the SM right-handed down-type quarks, which Yukawa
couple to the SM left-handed quark doublets $Q_L^i$ as
\beq
{\cal  L }_{\rm NP}^{\rm (V)} = {\bar D} (i\slashed{D} - M) D
  - (\lambda_i {\bar D_R} H^{\dagger} Q_L^i + {\rm h.c.} ) \,,
\eeq
These new interactions generate $Z$ couplings, e.g., in this Model~V to the
quarks,
\beq\label{bsmZ}
{\cal L}_Z^{\rm (V)} =
  - \sum_{i,j} \left( {\lambda_i^* \lambda_j\, m_Z^2 \over g_Z\, M^2}\right) 
  {\bar d}^{i}_L \gamma^{\mu} d^{j}_L\, Z_\mu \,,
\eeq
which contribute to, and are constrained by, flavor-changing neutral currents.

\begin{figure}[tb]
\centerline{\includegraphics[width=.3\textwidth]{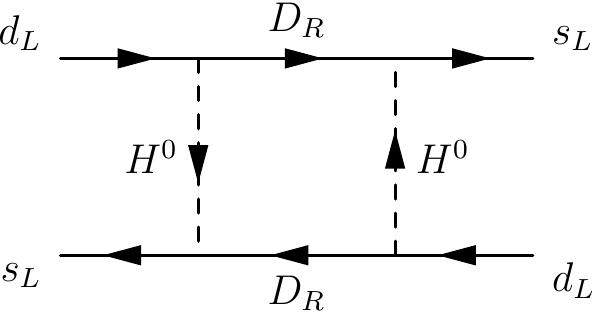} \hfil
\includegraphics[width=.3\textwidth]{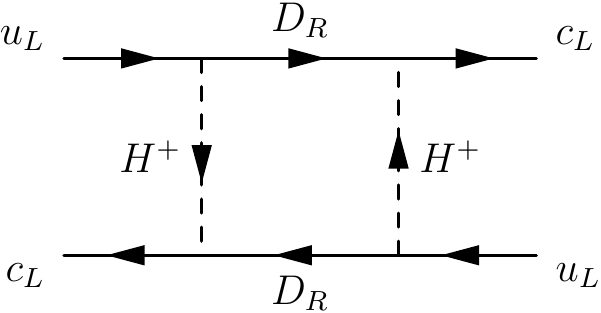}}
\caption{One-loop vector-like fermion contributions to $K$ and $D$ mixing in 
Model~V~\cite{Ishiwata:2015cga}.}
\label{fig:mixing}
\end{figure}

These models also generate dimension-6 four-fermion operators, which contribute
to neutral meson mixing.  At tree level, the $Z$ contribution in
Eq.~(\ref{bsmZ}) yields coefficients of the form $(\lambda_i \lambda_j^{*})^2
v^2/M^4$.  At one loop, coefficients of order $( \lambda_i \lambda_j^{*})^2 /(4
\pi M)^2$ are generated, which are neither CKM nor quark-mass suppressed,
seemingly not considered in the literature.  For large $M$, these one-loop
contributions are more important than tree-level $Z$ exchange.  They are
independent of the Higgs vacuum expectation value, $v$, and arise from short
distances $\sim 1/M$.  They can be calculated in the symmetric phase from the
box diagrams in Fig.~\ref{fig:mixing} with virtual scalars and the heavy
vector-like fermions.  The resulting effective Lagrangian in Model~V
is~\cite{Ishiwata:2015cga},
\beq
{\cal L}_{\rm mix}^{\rm (V)} = - {(\lambda_i^* \lambda_j )^2\over 128 \pi^2 M^2}
  \bigg[ \sum_{klmn} \big( {\bar u}_L^k V_{ki}\, \gamma_{\mu} V^{\dagger}_{jl}\,
  u_L^l \big) 
\big( {\bar u}_L^m V_{mi}\,\gamma_{\mu} V^{\dagger}_{jn}u_L^n \big)
  + \big( {\bar d}_L^i \gamma_{\mu}d_L^j \big)
  \big( {\bar d}_L^i \gamma^{\mu}d_L^j \big) \bigg] + {\rm h.c.}
\eeq

Table~\ref{tab:bounds} shows the bounds on 5 of the 11 models for illustration
(see also Ref.~\cite{Bobeth:2016llm}).  The upper rows for each model contain
the current bounds, and the lower rows show the expected sensitivities in the
next generation of experiments (in the next decade or so).  For the vector-like
fermions that couple to SM quarks, the bounds are  shown separately from $\Delta
F=1$ and $\Delta F=2$ processes.  For the $\Delta F=2$ bounds on the 1--2
generation couplings, the bounds are shown separately on the real and imaginary
parts, since $\epsilon_K$ probes much higher scales than $\Delta m_K$ in these
models.  (In the other cases the differences are of order unity.)  We learn that
the next generation of experiments will improve the mass scale sensitivities in
the leptonic (hadronic) models by up to a factor of $\sim 7$ ($\sim 4$).

\begin{table}[tb] \tabcolsep 6pt
\centerline{\begin{tabular}{c@{\extracolsep{2pt}}c|cc|cc|cc}
\hline\hline
\multirow{2}{*}{Model} &  \multicolumn{1}{c|}{Quantum}  &
  \multicolumn{6}{c}{Bounds on $M/\TeV$ and $\lambda_i \lambda_j$
  for each $ij$ pair}  \\
& \multicolumn{1}{c|}{numbers} & \multicolumn{2}{c|}{$ij=12$} 
  & \multicolumn{2}{c|}{$ij=13$} & \multicolumn{2}{c}{$ij=23$}  \\
\hline
II & $(1,3,-1)$ & \multicolumn{2}{c|}{220$^a$} & \multicolumn{2}{c|}{4.9$^b$}
 & \multicolumn{2}{c}{5.2$^c$}
\\
 &  & \multicolumn{2}{c|}{1400$^a$} & \multicolumn{2}{c|}{13$^b$}
  & \multicolumn{2}{c}{15$^c$}
\\
III & $(1,2,-1/2)$ & \multicolumn{2}{c|}{310$^a$} & \multicolumn{2}{c|}{7.0$^b$}
  & \multicolumn{2}{c}{7.4$^c$}
\\
 &  & \multicolumn{2}{c|}{2000$^a$} & \multicolumn{2}{c|}{19$^b$}
  & \multicolumn{2}{c}{21$^c$} 
\\ \cline{3-8}
& & $\Delta F=1$ & $\Delta F=2$ & $\Delta F=1$ & $\Delta F=2$
  & $\Delta F=1$ & $\Delta F=2$
\\ \cline{3-8}
V & $(3,1,-1/3)$ & 66$^d$ [100]$^e$ & \{42, 670\}$^f$
  & 30$^g$ & 25$^h$ & 21$^i$ & 6.4$^j$
\\
 &  & 280$^d$ & \{$100$, 1000\}$^f$  & 60$^l$ & 61$^h$ & 39$^k$ & 14$^j$
\\
VII & $(3,3,-1/3)$  & 47$^d$ [71]$^e$ & \{47, 750\}$^f$ 
  & 21$^g$ & 28$^h$ & 15$^i$ & 7.2$^j$
\\
 &  & 200$^d$ & \{$110$, 1100\}$^f$  & 42$^l$ & 68$^h$ & 28$^k$ & 16$^j$
\\
XI & $(3,2,-5/6)$  & 66$^d$ [100]$^e$ & \{42, 670\}$^f$
  & 30$^g$ & 25$^h$ & 18$^k$ & 6.4$^j$
\\
 &  & 280$^d$ & \{$100$, 1000\}$^f$  & 60$^l$ & 61$^h$ & 39$^k$ & 14$^j$
\\
\hline\hline
\end{tabular}}
\caption{Bounds in some of the vector-like fermion
models~\cite{Ishiwata:2015cga} on $M\, [\TeV] /\sqrt{|\lambda_i \lambda_j|}$ in
the leptonic models, and from the $\Delta F=1$ constraints on the hadronic
models.  The $\Delta F=2$ bounds show $M/\sqrt{|\lambda_i\lambda_j|^2}$, except
for $K$ meson mixing we show $\big\{ M/\sqrt{|{\rm
Re}(\lambda_i\lambda_j^*)^2|}, \ M/\sqrt{|{\rm Im}(\lambda_i\lambda_j^*)^2|}\,
\big\}$.  The strongest bounds arise, or are expected to arise, from: $a)$ $\mu$
to $e$ conversion; $b)$ $\tau\to e\pi$; $c)$ $\tau\to \mu\rho$; $d)$
$K\to\pi\nu\bar\nu$; $e)$ $K_L\to\mu^+\mu^-$ (this involves $|{\rm
Re}(\lambda_1\lambda_2^*)|$ and is in square brackets because prospects for
improvements are weak); $f)$ $K$ mixing; $g)$ $B\to\pi\mu^+\mu^-$; $h)$ $B_d$
mixing; $i)$ $B\to X_s\ell^+\ell^-$; $j)$ $B_s$ mixing;  $k)$
$B_s\to\mu^+\mu^-$, $l)$ $B_d\to\mu^+\mu^-$.}
\label{tab:bounds}
\end{table}

\subsection{Top, higgs, and new physics flavor}

These are vast topics which I could not cover in detail in the talk, nor is it
possible here.  

Top quarks in the SM decay almost exclusively to $bW$, with the second largest
branching fraction ${\cal B}(t\to sW) < 2\times 10^{-3}$.  Particularly clean
probes of the SM are FCNC top decays, for which the SM predictions are below the
$10^{-12}$ level.  The current bounds are roughly at the level ${\cal B}(t\to
qZ) \lsim 10^{-3}$, ${\cal B}(t\to qg) \lsim 10^{-4}$, and ${\cal B}(t\to qh)
\lsim 0.5\%$, with the precise limits depending on the ratio of $q=u,c$ produced
by new physics.  The ultimate LHC sensitivities are expected to be
about a factor of $10^2$ better, hence any observation would be a clear sign of
NP.  There is obvious complementarity between FCNC searches in the top sector,
and low energy flavor physics bounds.  Since $t_L$ is in the same $SU(2)$
doublet as $b_L$, several operators have correlated effects in $t$ and $b$
decays.  For some operators, mainly those involving left-handed quark fields,
the low energy constraints exclude a detectable LHC signal, whereas other
operators are still allowed to have large enough coefficients to yield
detectable NP signals at the LHC (see, e.g., Ref.~\cite{Fox:2007in}).

The experimental richness of higgs physics, that several production mechanisms
and many decay channels can be probed, are to a large extent due to the
particular values of the Yukawa couplings. The quark and lepton couplings, and
$Y_t$ in particular, are important for higgs decays, as well as to determine the
production cross sections from $gg$ fusion, higgs-strahlung, $t\bar t$ and $WZ$
fusion.  The LHC has (almost) measured the $h\tau^+\tau^-$ coupling, and will
also determine $h\mu^+\mu^-$ and $hb\bar b$, if they are near their SM values. 
Should the LHC or another future collider detect deviations from the SM
branching ratios or observe flavor-non-diagonal higgs decays, that would of
course be incredibly significant (for a recent discussion, see, e.g.,
Ref.~\cite{Nir:2016zkd}).

Any new particle that couples to the quarks and/or leptons, potentially
introduces new flavor violating parameters.  For example, in low energy
supersymmetry, which is the favorite NP scenario of a large part of our
community, squark and slepton couplings may yield measurable effects in FCNC
processes and $CP$ violation, give rise to detectable charged lepton flavor
violation (CLFV), such as $\mu\to e \gamma$, etc.  Observable $CP$ violation is
then also possible in neutral currents and electric dipole moments, for which
the SM predictions are below the near future experimental sensitivities.  The
supersymmetric flavor problems, that TeV-scale SUSY models with generic
parameters are excluded by FCNC and $CP$ violation measurements, can be
alleviated in several scenarios: (i) universal squark masses, when $\Delta\tilde
m_{\tilde Q,\tilde D}^2 \ll \tilde m^2$ (e.g., gauge mediation); (ii) alignment,
when $(K^d_{L,R})_{12} \ll 1$ (e.g., horizontal symmetry); (iii) heavy squarks,
when $\tilde m \gg 1\,\TeV$ (e.g., split SUSY).  All viable models incorporate
some of these ingredients.  Conversely, if SUSY is discovered, mapping out its
flavor structure may help answer questions about even higher scales, e.g., the
mechanism of SUSY breaking, how it is communicated to the MSSM, etc.

An important implication of flavor constraints for SUSY searches is that the LHC
bounds are sensitive to the level of (non-)degeneracy assumed.  Most SUSY
searches assume that the first two generation squarks, $\tilde u_{L,R}$, $\tilde
d_{L,R}$, $\tilde s_{L,R}$, $\tilde c_{L,R}$, are all degenerate, which
increases signal cross sections.  Relaxing this assumption consistent with
flavor bounds, results in substantially weaker squark mass limits from the LHC
Run~1, around the 500\,GeV scale~\cite{Mahbubani:2012qq}.  Thus, there is a
tight interplay between the flavor physics and LHC high-$p_T$ searches for new
physics.  If there is new physics at the TeV scale, its flavor structure must be
highly non-generic to satisfy current bounds, and measuring small deviations
from the SM in the flavor sector would give a lot of information complementary
to ATLAS \& CMS.  The higher the scale of new physics, the less severe the
flavor constraints are.  If NP is beyond the reach of the LHC, flavor physics
experiments may still observe robust deviations from the SM, which would point
to an upper bound on the next scale to probe.

\section{Final comments and ultimate sensitivity}
\label{sec:concl}

The main points I tried to convey through some examples were:

\begin{itemize}\vspace*{-6pt}\itemsep -2pt

\item $CP$ violation and FCNCs are sensitive probes of short-distance physics
in the SM and for~NP;

\item Flavor physics probes energy scales $\gg\!1\,\TeV$, the sensitivity
limited by statistics, not theory;

\item For most FCNC processes NP\,/\,SM $\gsim 20\%$ is still allowed, so there
is plenty of room for NP;

\item Of the several tensions between data and SM predictions, some may soon
become definitive;

\item Precision tests of SM will improve by $10^1$\,--\,$10^4$ in many channels
(including CLFV);

\item There are many interesting theory problems, relevant for improving
experimental sensitivity;

\item Future data will teach us more about physics at shorter distances,
whether NP is seen or not, and could point to the next energy scale to explore.

\end{itemize}\vspace*{-6pt}

With several new experiments starting (NA62, KOTO, Belle~II, mu2e, COMET, etc.)
and the upcoming upgrade of LHCb, the flood of new data will be fun and exciting
(see Refs.~\cite{Bediaga:2012py, Belle2predictions} for reviews of planned
flavor experiments and their sensitivities).  It will allow new type of
measurements, and more elaborate theoretical methods to be used and tested.  The
upcoming experiments also challenge theory, to improve predictions and to allow
more measurements to probe short distance physics with robust discovery
potential.  Except for the few cleanest cases, improvements on both sides are
needed to fully exploit the future data sets.  I am optimistic, as order of
magnitude increases in data always triggered new theory developments, too.

It is also interesting to try to estimate the largest flavor physics data sets
which would be useful to increase sensitivity to new physics, without being
limited by theory uncertainties.\footnote{In measurements without SM
backgrounds, such as setting bounds on $\mu\to e$ conversion or $\tau\to 3\mu$
decay, the mass-scale sensitivity (to a dimension-6 NP operator) scales like
$\Lambda \propto (\mbox{bound})^{-1/4}$.  In measurements constraining SM--NP
interference, $\Lambda \propto (\mbox{uncertainty})^{-1/2}$, and at some point
precise knowledge of the SM contribution becomes critical.}  For charged lepton
flavor violation, the SM predictions (from penguin and box diagrams with
neutrinos) are (tens of) orders of magnitudes below any foreseeable experimental
sensitivity, so if technology allows significant improvements, I think the
justification is obvious (as it is for electric dipole moment searches).  In
quark flavor physics the situation is more complex.  Amusingly, even in 2030,
there will be theoretically clean $B$ decay modes in which (experimental
bound)\big/SM $\gtrsim 10^3$, e.g., $B\to \tau^+\tau^-$, $B\to e^+e^-$, and
probably some more.  However, based on what is known today, some observables
will become limited by theory (hadronic) uncertainties.  Identifying how far NP
sensitivity can be extended is interesting, at least in principle, so below is a
list for which 50/ab Belle~II and 50/fb LHCb data will not even come within an
order of magnitude of the ultimately achievable sensitivities.  Of course, on
the relevant time scale lots of progress will take place (see
Ref.~\cite{Butler:2013kdw} for estimates of future lattice QCD uncertainties)
and new breakthroughs are also possible.  

\begin{figure}[t]
\centerline{\includegraphics[width=.6\textwidth]{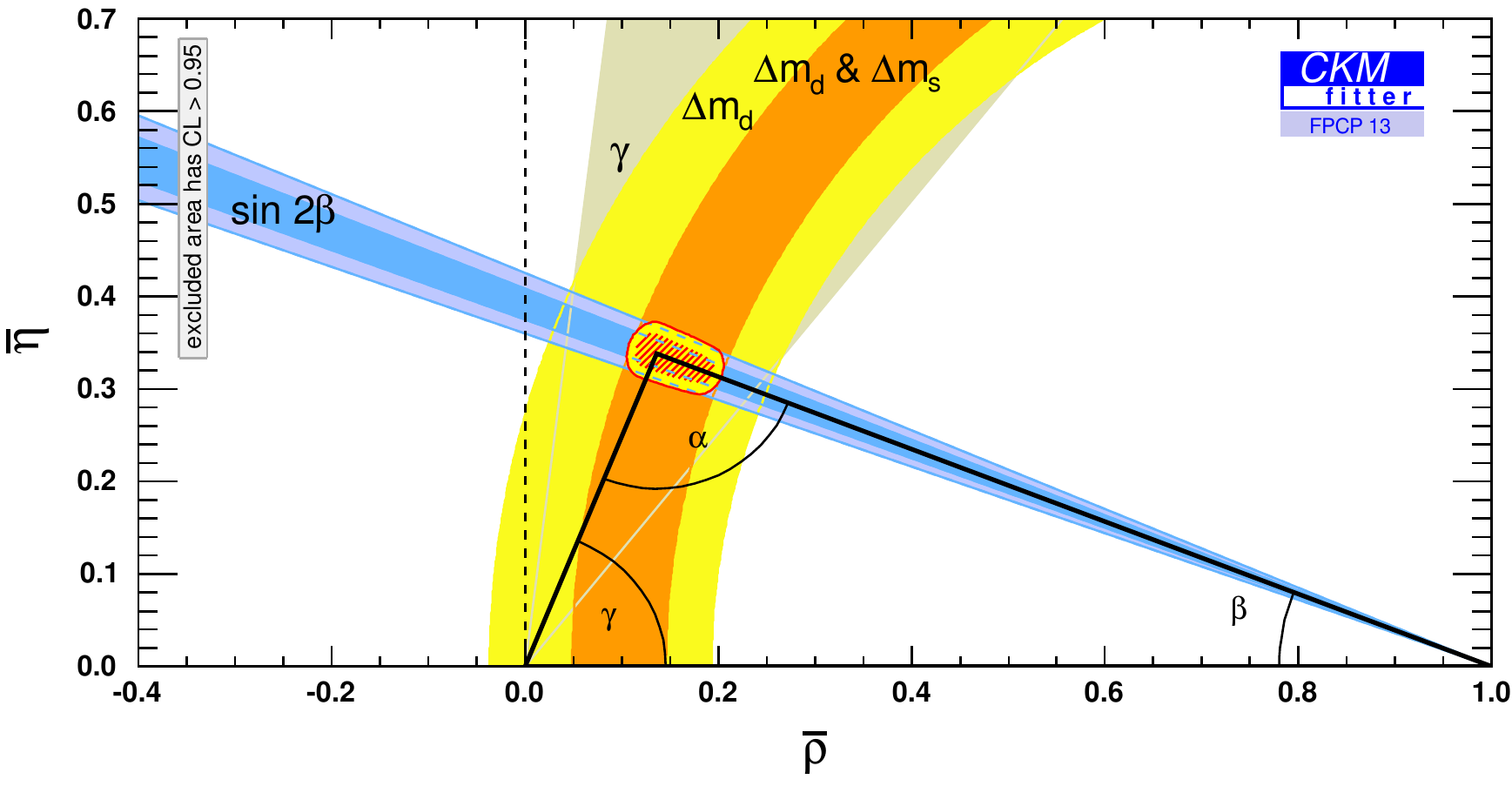}}
\caption{A test of the SM flavor sector that can improve by a factor of 10.}
\label{fig:unlimited}
\end{figure}

\begin{itemize}\vspace*{-6pt}\itemsep -2pt

\item Probably the theoretically cleanest observable in the quark sector is the
determination of the CKM phase $\gamma$ from tree-level $B$ decays.  Irreducible
theory uncertainty only arises from higher order weak
interaction~\cite{gammaTH}.  So the main challenges are on the experimental
side.  

\item The theory uncertainty for the semileptonic $CP$ asymmetries, $a_{\rm
SL}^{d,s}$, discussed is Sec.~\ref{sec:status} and in Fig~\ref{fig:cpvmix}, are
also much below~\cite{Laplace:2002ik, Lenz:2011ti} the expected 50/ab Belle~II
and 50/fb LHCb sensitivities.

\item Another set of key observables are $B_{s,d}\to\mu\mu$ and $B\to\ell\nu$,
where the nonperturbative theory inputs are only the decay constants, which will
soon be known with $<1\%$ uncertainties.  In contrast, the expectation for the
accuracy of $B_d\to\mu\mu$ with the full LHC data is ${\cal O}(20\%)$.

\item It is often stated that the determination of $|V_{ub}|$ is theory
limited.  This entirely depends on the measurements available.  In principle,
the theoretically cleanest $|V_{ub}|$ determination I know, which only uses
isospin, would be from ${\cal B}(B_u\to \ell\bar\nu) / {\cal B}(B_d\to
\mu^+\mu^-)$~\cite{BGckm06}.

\item I think that the SM prediction for $CP$ violation in $D^0$\,--\,$\D0bar$
mixing is below the expected sensitivities on LHCb and Belle~II.  To establish
this robustly, however, more theory work is needed (especially given the recent
history of hints of $CP$ violation in $D$ decay).

\item For $K^+\to\pi^+\nu\bar\nu$ and especially for $K_L\to\pi^0\nu\bar\nu$,
the current plans for NA62 and KOTO will stop short of reaching the ultimate
sensitivity to NP achievable in these decays.

\end{itemize}\vspace*{-6pt}

Thus, I guess(timate) that $\sim 100$ times the currently envisioned 50/ab
Belle~II and 50/fb LHCb data sets would definitely allow for the sensitivity to
short distance physics to improve.  Whether any of these ultimate sensitivities
can be achieved at a tera-$Z$ machine, an $e^+e^-$ collider running on the
$\Upsilon(4S)$, or utilizing more of the LHC's or/and a future hadron collider's
full luminosity, is something I hope we shall soon have even more compelling
reasons to seriously explore.

\medskip
\paragraph{Acknowledgments}
I thank the organizers of the CKM2016 conference for the invitation to an
exciting and fun meeting.
I thank Marat Freytsis, Tim Gershon, Koji Ishiwata, Michele Papucci, Dean
Robinson, Josh Ruderman, Filippo Sala, Karim Trabelsi, Phill Urquijo, and Mark
Wise, for recent collaborations and/or discussions that shaped the views
expressed in this talk. This work was supported in part by the  U.S.\ Department
of Energy under contract DE-AC02-05CH11231.

\end{document}